\documentclass[epj]{svjour}
\usepackage{graphics}
\usepackage{enumitem}
\usepackage[colorlinks]{hyperref}
\def\gsimeq{\,\,\raise0.14em\hbox{$>$}\kern-0.76em\lower0.28em\hbox {$\sim$}\,\,}
\def\lsimeq{\,\,\raise0.14em\hbox{$<$}\kern-0.76em\lower0.28em\hbox{$\sim$}\,\,}
\begin{document}
\newcommand{\SG}[1]{\textcolor{blue}{#1}}

\title {Shell-model based study of the direct capture in neutron-rich nuclei}
\author{K. Sieja\inst{1} and S. Goriely\inst{2}}
\institute{1. Universit\'e de Strasbourg, IPHC, 23 rue du Loess 67037 Strasbourg, France 
CNRS, UMR7178, 67037 Strasbourg, France\\
2. Institut d'Astronomie et d'Astrophysique, Universit\'e Libre de Bruxelles, CP-226, 1050 Brussels, Belgium}

\abstract{
The radiative neutron capture rates for isotopes of astrophysical interest are commonly calculated within
the statistical Hauser-Feshbach reaction model. Such an approach, assuming a high level density in the compound system, 
can be questioned in light and neutron-rich
nuclei for which only a few or no resonant states are available. Therefore, in this work we focus on 
the direct neutron-capture process.
We employ a shell-model approach in several model spaces with well-established effective interactions
to calculate spectra and spectroscopic factors in a set of 50 neutron-rich target nuclei in different mass regions, including
doubly-, semi-magic and deformed ones. Those theoretical energies and spectroscopic 
factors are used to evaluate direct neutron capture rates
and to test global theoretical models using average spectroscopic factors and level densities based on the Hartree-Fock-Bogoliubov plus combinatorial
method. The comparison of shell-model and global model results reveals several discrepancies 
that can be related to problems in level densities.
All the results show however that the direct capture is non-negligible with respect to the by-default Hauser-Feshbach
predictions and can be even 100 times more important for the most neutron-rich nuclei close to the neutron drip line. 
}

\maketitle

\section{Introduction}
Nuclear reactions of astrophysical interest often concern unstable or even exotic species for which no experimental data exist. Although significant efforts have been devoted in the past decades, experimental information only covers a minute fraction of the entire data set required for nuclear astrophysics. Moreover, the energy range for which experimental data is available is restricted to the small range that can be studied by present experimental setups. In that case, only theoretical predictions can fill the gaps. One of this specific examples concerns the rapid neutron-capture process (or r-process) called for to explain the origin of about half of the elements heavier than iron observed in nature (for a review, see \cite{Arnould07,Arnould20}). The r-process is believed to take place in environments characterized by the high neutron densities, such that successive neutron captures can proceed into neutron-rich regions well off the $\beta$-stability valley. It  involves a large number (typically five thousands) of unstable nuclei for which many different properties have to be determined and cannot be obtained experimentally. One of such fundamental properties concern the radiative neutron capture reaction.

So far, the neutron capture rates are usually evaluated within the statistical Hauser-Feshbach (HF) model \cite{Goriely08}. The model makes the fundamental assumption that the capture process takes place with the intermediary formation of a compound nucleus in thermodynamic equilibrium. The energy of the incident particle is then shared more or less uniformly by all the nucleons before releasing the energy by particle emission or $\gamma$-de-excitation.
The formation of a compound nucleus is usually justified by assuming that the nuclear level density (NLD) in the compound system at the projectile incident energy is large enough to ensure an average statistical continuum superposition of available resonances \cite{Satchler80}. However, when the number of available states in the compound system is relatively small, the validity of the HF predictions has to be questioned, the neutron capture process being possibly dominated by direct electromagnetic transitions to a bound final state rather than through a compound intermediary. The direct capture (DC) proceeds via the excitation of only a few degrees of freedom on much shorter time scale reflecting the time taken by the projectile to traverse the target. For the DC process, the mean free path of the incident particle is comparable with the size of the nucleus and the particle ejection occurs preferentially at forward angles. It has become clear, however, that the DC process is important, and often dominating at the very low energies of astrophysical interest, especially for light or exotic nuclei systems for which few, or even no resonant states are available \cite{Oberhummer91,Mengoni95,Beer96,Descouvemont03,Descouvemont08,Xu2012,Xu14}. This should be also the case for medium-mass or heavy neutron rich nuclei produced by the r-process nucleosynthesis. However, all existing simulations are up to now exclusively relying on radiative neutron capture cross sections obtained within the HF approach.

In the present work we compute energy levels and spectroscopic factors of nuclei far from stability 
within the large-scale shell model approach
and use them to determine direct neutron-capture cross sections within the potential model. The theoretical framework is defined in Sec.~\ref{sec:th}. The shell-model prediction of the DC component is discussed in Sec.~\ref{sec:smpred} and compared with global microscopic models in Sec.~\ref{sec:global}. 
Predictions based on the shell model results are compared to those based on global theoretical models employing
NLD from the Hartree-Fock-Bogoliubov (HFB) plus combinatorial method. The direct capture cross    
sections obtained with the shell-model input are also confronted with the usual HF
estimates of the neutron capture.

\section{Theoretical framework}
\label{sec:th}
\subsection{Direct-capture calculations}
\label{sec:th1}
The calculations of the DC cross sections are done within the potential model, following the method described in \cite{Xu2012}.
The potential model is employed to study the neutron DC reaction describing 
the transition from the initial scattering state $A+n$ directly to the final nucleus $B$ with accompanying $\gamma$-ray emission. 
The allowed electric dipole ($E1$), electric quadrupole ($E2$) and magnetic dipole ($M1$) transitions to the ground 
state as well as all possible excited states in the final nucleus are taken into account.

The neutron DC cross section for $A(n,\gamma)B$ can be expressed as \cite{Goriely1997}
\begin{eqnarray}
&&\sigma^{DC}(E)=\sum_{f=0}^{E^{x}_{B}}S_f\sigma_{dis}(E)\nonumber\\
&+&\langle S_f\rangle \int_{E^{x}_{B}}^{S_n}\sum_{J_f,\pi_f}\rho(E_f,J_f,\pi_f)\times\sigma_f^{cont}dE_f\,.
\label{eqpotall}
\end{eqnarray}
\noindent
Below $E^{x}_{B}$, the sum runs over all the available discrete final states in the residual nucleus $B$, 
which are usually experimental levels if available.
$S_f$ is the spectroscopic factor describing the overlap between the antisymmetrized wave function of the initial system $A+n$ and the final state $B^{x}$. 
Above $E^{x}_{B}$, the summation is replaced by a continuous integration over a spin- and parity-dependent NLD ($\rho(E_f,S_f,\pi_f)$)
and the spectroscopic factor by an average quantity $\langle S_f \rangle$.
In the present study in neutron-rich nuclei we employ theoretically deduced levels and spectroscopic factors
from 3 different models, namely
\begin{enumerate}[label=(\roman*)]
\item the shell-model discrete energy levels with corresponding spectroscopic factors, computed as described in the next section, which we refer to as SM.
\item combinatorial NLD from Ref. \cite{Goriely-LD} with an average value of the spectroscopic factor $\langle S_f\rangle=0.5$, which we refer to as
intrinsic model.
\item one-particle one -hole neutron excitations deduced from the combinatorial NLD above
with an average spectroscopic factor $\langle S_f\rangle=1.0$, which we will refer to as 1p-1h model.
\end{enumerate}
For models (ii) and (iii), the cross section is evaluated using the second part of Eq. \ref{eqpotall} with the corresponding average value
of $S_f$ equal to 0.5 or 1, respectively, except for the ground state for which $S_f$= 0.347, a value deduced from
the compilation of all the known spectroscopic factors \cite{Goriely98}. The values of $\langle S_f\rangle$ were shown to minimize
deviations between theory and experiment in nuclei close to the valley of $\beta$-stability where both models yield similar results \cite{Xu2012}.
In the shell-model case we compute all the necessary discrete levels with their corresponding $S_f$ values, thus the first part of Eq. \ref{eqpotall} is used
to evaluate the cross section. Also note that to evaluate the predictive power of models (ii) and (iii), no experimental information on excited levels is included in the calculation of the DC cross section.

The potential model calculates the transition matrix elements between the initial and the 
final states by sandwiching the electromagnetic operators in the long wave-length limit. 
It is usually enough to consider the $E1$, $E2$ and $M1$ transitions which is the case
of the present work. The complete set of equations used to calculate 
the matrix elements of the electromagnetic operators can be found in Ref.~\cite{Xu2012}.
The radial wave functions are obtained from the solution of the 2-body Schr\"odinger equation
with a central potential; in our case the Koning-Delaroche potential \cite{KD}  was employed in all calculations.
Since our objective is to test the DC values based on global prescriptions for NLD and $\langle S_f\rangle $ 
against those based on SM far from stability, all 
nuclear ingredients except excitation spectra and spectroscopic factors are kept the same in
the reaction calculations. Experimental masses are used whenever available, otherwise predictions from
HFB-27 mass model based on the 
BSk27 Skyrme interaction are used \cite{Goriely-HFB27}. 

\subsection{Shell-model calculations of energy spectra and spectroscopic factors}
\label{sec:th2}
The shell-model calculations were performed using the Strasbourg m-scheme code ANTOINE developed by E. Caurier \cite{ANTOINE,RMP},
which permits the use of large configuration spaces when necessary. 
The relevant spectroscopic factors can be obtained by calculating the response function for the neutron-creating 
operator acting on the shell-model ground-state wave function of the target nucleus. We have calculated this response function 
using the Lanczos strength function method with 100 iterations.
Such a procedure allows to get a convergence of at least 10 lowest states of each spin,
which exhausts the number of levels located below the $S_n$ value in most neutron-rich nuclei.
Otherwise, at higher excitation energies the approximation allows us to obtain the 
spectroscopic strengths per excitation energy interval, which is directly comparable
to the prescription used in Ref.~\cite{Xu2012}. 
We evaluated this way excited states and spectroscopic
factors starting from the ground state of 50 neutron-rich targets, both even-even and even-odd, 
spherical and well deformed.
The calculations were performed in various model spaces with the following interactions:

\begin{itemize}
\item{SDPF-U}\\
Full space diagonalizations in the proton $sd$-  neutron $pf$-shell 
were performed using the well-established SDPF-U interaction \cite{SDPF}
for argon and sulfur isotopes with $N=24-32$. 
The SDPF-U interaction was designed
to describe the neutron-rich nuclei around $N=28$ in a $0\hbar\omega$ space,
therefore, it is applicable to nuclei with $8\le Z\le 20$ and $20\le N \le 40$, covering perfectly
the region chosen for this study. The interaction was first used to describe the vanishing
of the $N=28$ shell-closure below $^{48}$Ca and predicted correctly the deformation of $^{42}$Si \cite{Bastin2007}
and the spectroscopic factors of $^{46}$Ar \cite{Gaudefroy2006}.
Since its publication it was frequently applied to this region of nuclei with a great success, for example
it reproduces well the triple shape-coexistence in neutron-rich sulfurs \cite{Gaudefroy2009,Force-S44,Santiago-Gonzalez},
and many other spectroscopic properties in Ar and S isotopes, see e.g. Refs.~\cite{Liu-Ar52,Gade-S45,Calinescu-Ar46,Saxena,Chevrier}.

\item{LNPS}\\
The LNPS interaction \cite{Lenzi2010} defined in the proton $pf$- neutron $pf{9/2}d_{5/2}$ space was 
first introduced to describe the onset of deformation below $N=40$, known as the second island of inversion.
In addition to the reproduction of the known data \cite{Lenzi2010,Ljungvall}, it predicted the range
of this phenomenon which triggered a whole experimental campaign. The LNPS was since then used 
in dozens of experimental interpretations of detailed nuclear structure in the region, see e.g.
Refs.~\cite{Gade-Ti60,Modamio,Diriken,Georgi2011,Elisa2011,Eda,Vajta,Dijon,Co-76,Meisel,Shand}.
In particular, it is worth noticing its success in the study of the spectroscopic factors, see for example
Refs. \cite{Morfouace2015,Giron,Morfouace2016,Elekes,Orlandi}. 
Here we employed the interaction to compute spectroscopic factors in spherical nickel and prolate-deformed chromium chains from $N=36$ to $N=52$.
The calculations were performed with 8p-8h truncation with respect to the $Z=28$ and $N=40$ gaps. 

Since in the neutron-rich nuclei around $N=50$, the $d_{5/2}$ and $s_{1/2}$
orbitals are possibly degenerate (in $^{79}$Zn experiment suggests $1/2^+$ at 1.1~MeV
with $S_f=0.41(10)$\cite{Orlandi}), it is necessary to unblock the possibility of
the neutron capture on the $s_{1/2}$ and higher positive parity orbitals close to $N=50$. Therefore, for the
nuclei closest to the $N=50$ and passed the shell closure, we considered for comparison
a larger valence space, described in the next point.
\item{LNPS-GDS}\\
We developed an interaction comprising $pf$ shell for protons and 
$gds$ shell for neutrons starting from a realistic $V_{lowk}$ potential based on
the N3LO interaction \cite{N3LO,vlowk}. The monopoles of the $pfgd$ part were replaced by those
of the LNPS, therefore we dubbed this new interaction LNPS-GDS.
The single-particle energies were fixed to give the same spectrum of $^{79}$Ni as in
the effective interaction NI78-II \cite{Litzinger,Czerwinski-Br}.
Monopole adjustements were done to preserve the gaps in $^{78}$Ni
and the physics of nuclei around $N=50$, {\it i.e.} energies and transition rates of $^{80}$Zn \cite{Zn80}, 
spectra of $^{79}$Zn \cite{Orlandi} and
the yrast bands of the heavy nickel isotopes \cite{Elekes,NNDC,Ni78-nature}.
With such an interaction calculations in the $pf$ space for protons and $gds$ space for neutrons
were performed, with up to 6p-6h excitations across the $Z=28$ and $N=50$ gaps for the nickel and chromium chains,
from $N=42$ to $N=52$. 
\item{NI78-II}\\
Finally, we examined several Ge and Se isotopes above the $N=50$ closure.
They can be described in the proton $f_{5/2}p_{3/2}p_{1/2}g_{9/2}$ and neutron
$d_{5/2}d_{3/2}s_{1/2}g_{7/2}h_{11/2}$ 
model space with the so-called NI78-II interaction from
Refs.~\cite{Litzinger,Czerwinski-Br} which was optimized for neutron-rich nuclei up to $N=56$.
This interaction, used first for $Z=37-40$ isotopes \cite{sieja-Zr,Urban-y,Urban-Sr,Simpson-rb,urban-rb}, was shown successful 
to describe very detailed spectroscopy also in more exotic nuclei, towards the $^{78}$Ni core 
\cite{Shand,Czerwinski-Br,Czerwinski2019,Materna2015,Didierjean}.  
In particular, it reproduced correctly the known properties of $^{83,84,86,88}$Ge \cite{Kolos2013,Lettman} and $^{84-88}$Se 
nuclei \cite{Litzinger,Se87,Gratchev} and predicted non-axial degrees of freedom to be important in $N=52,54$ \cite{sieja2013}.
Here we chose to study those possibly $\gamma$-collective Ge and Se nuclei in the mass $84-89$ region. 
Full space diagonalizations of the Hamiltonian were achieved
in the valence space for cases considered in this work. 
\end{itemize}

\section{DC based on the shell-model predictions}
\label{sec:smpred}
\begin{center}
\begin{figure}
\resizebox{0.45\textwidth}{!}{\includegraphics{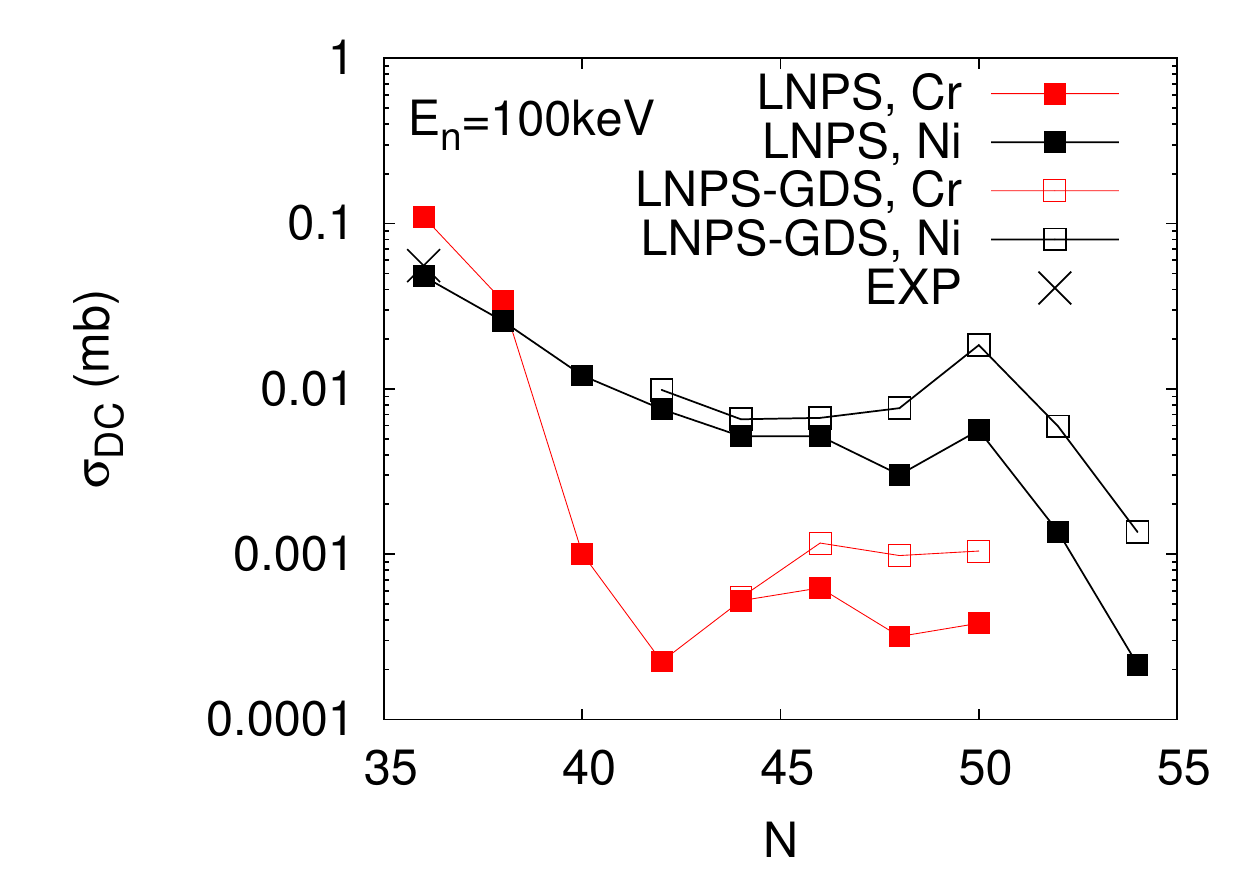}}
\caption{Neutron DC cross section for incident neutron energy $E_n=100$~keV
along the Cr and Ni chains. The results within LNPS model space
(filled symbols) are compared to those within the LNPS-GDS model space (open symbols).
EXP corresponds to the calculation of the cross section in $^{65}$Ni using the experimentally known
nuclear levels and spectroscopic factors from Ref.~\cite{NNDC}.}
\label{LNPSvsLNPS}
\end{figure}
\end{center}

The low-energy levels of the majority of nuclei studied in this work can be described within a model space comprising one
harmonic oscillator shell for each fluid, the only
exception being the nickel and chromium chains where the description of the deformation requires
the inclusion of at least the $d_{5/2}$ orbital already at $N=40$ and the influence 
of the $s_{1/2}$ orbital which increases towards $N=50$. As stated in the previous section, we used two different model spaces
dubbed here LNPS and LNPS-GDS. The impact of moving from one space to another is illustrated in Fig.~\ref{LNPSvsLNPS}
where the direct neutron capture cross section at the incident neutron energy $E_n=100$~keV obtained
within two model spaces in both chains are plotted. Shown is also the result of a calculation
in $^{65}$Ni which considers the energies and spectroscopic factors determined from experiment \cite{NNDC}. As seen,
the agreement between LNPS and experiment is very good as the present interaction reproduces 
very well detailed spectroscopic information in $^{65}$Ni and many other neighbouring nuclei,
including the available spectroscopic factors, see e.g. Refs.~\cite{Vajta,Morfouace2015,Giron,Orlandi}.  
\begin{figure}
\begin{center}
\resizebox{0.45\textwidth}{!}{\includegraphics{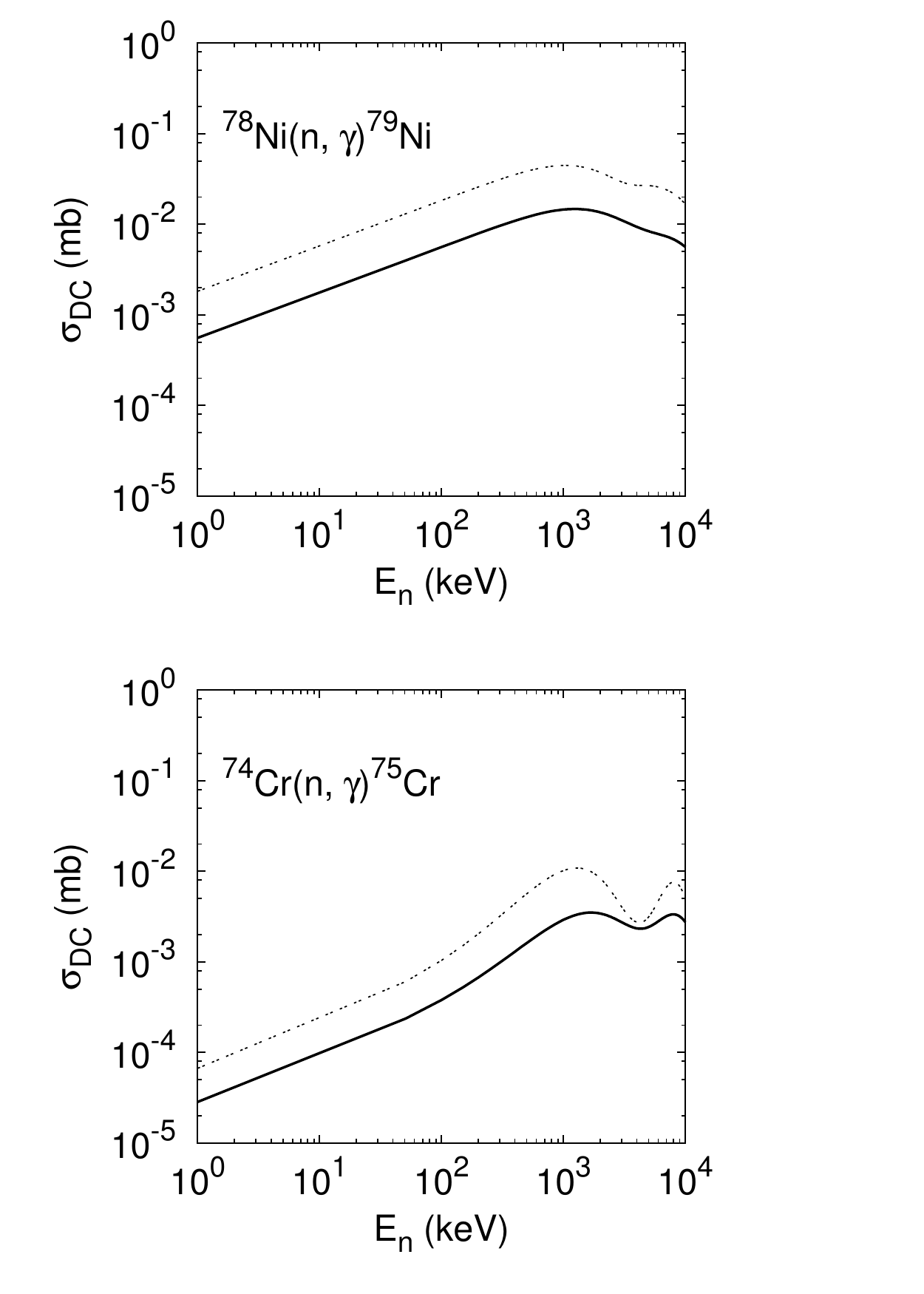}}
\caption{Neutron DC cross section at $N=50$ calculated in the LNPS model space
(solid line) and in the LNPS-GDS model space (dotted line).}
\label{nicr-dc}
\end{center}
\end{figure}

\begin{figure*}
\resizebox{0.95\textwidth}{!}{
\includegraphics{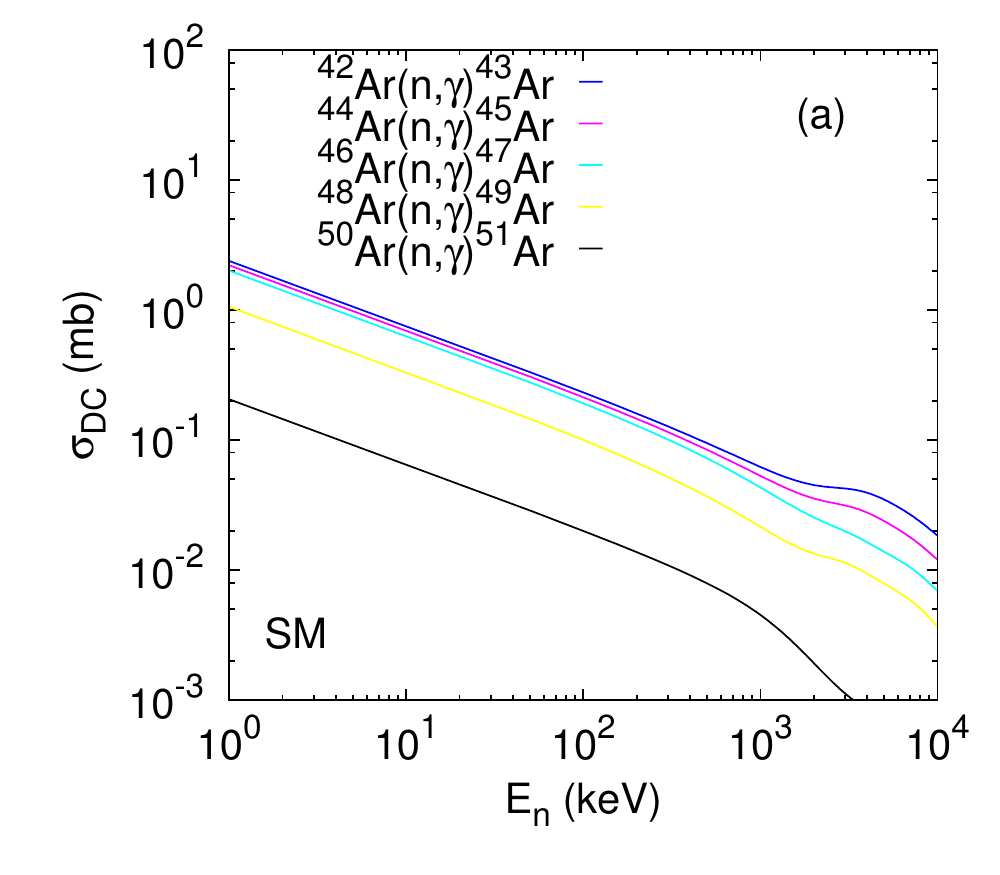}\includegraphics{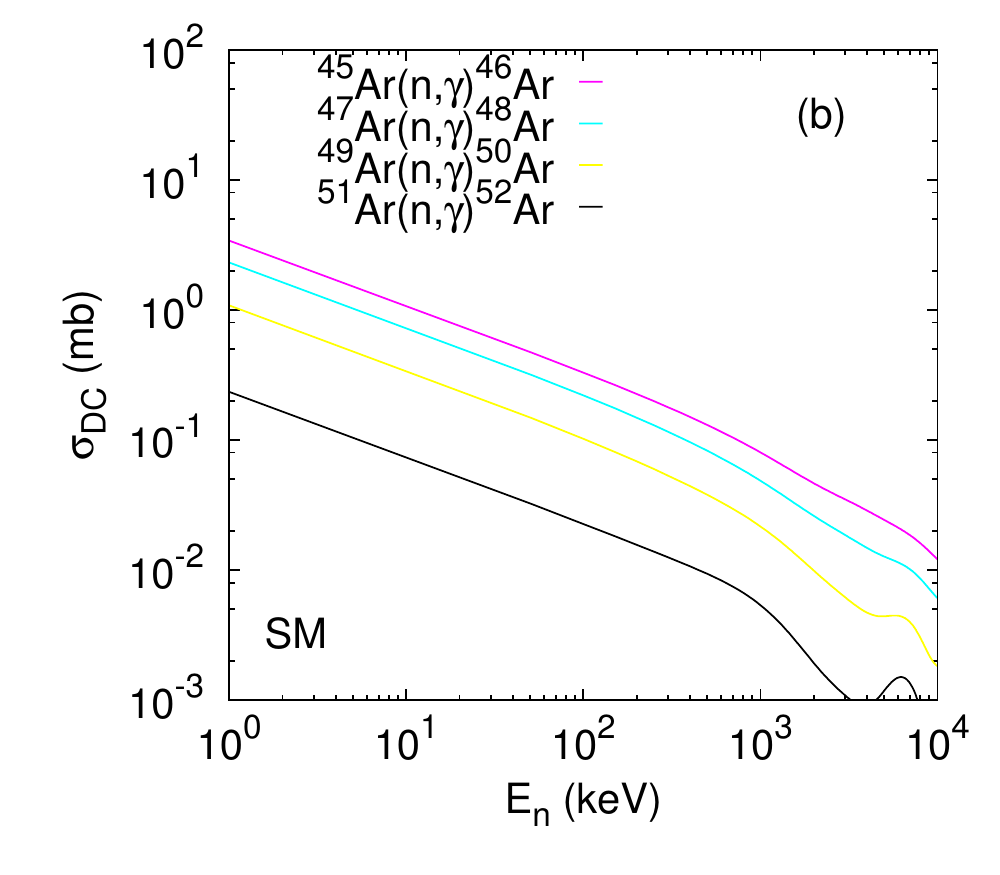}\includegraphics{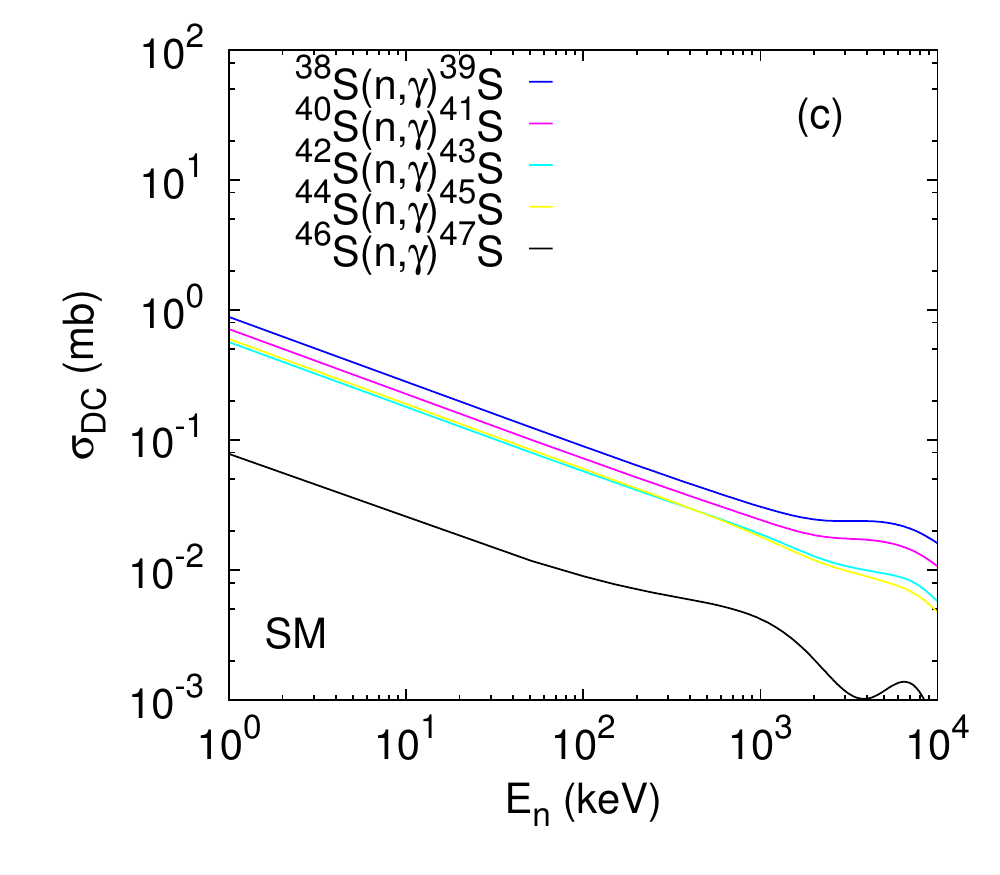}}
\end{figure*}
\begin{figure*}
\resizebox{0.95\textwidth}{!}{
\includegraphics{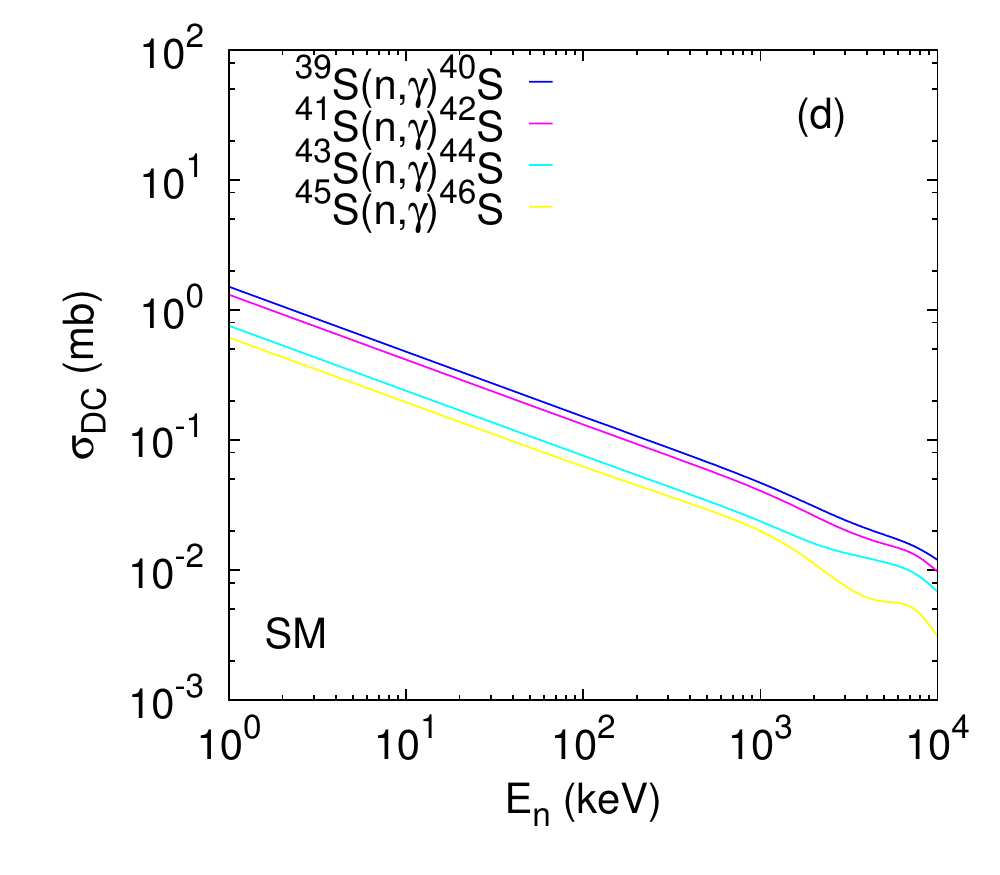}\includegraphics{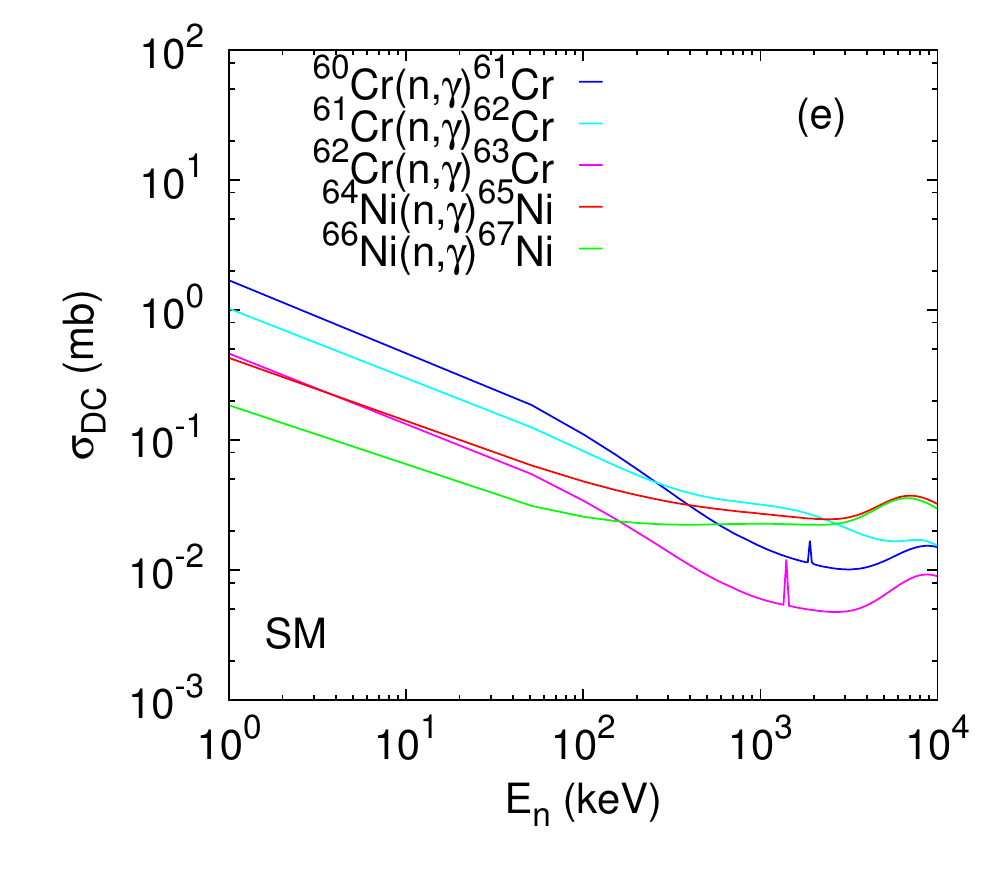}\includegraphics{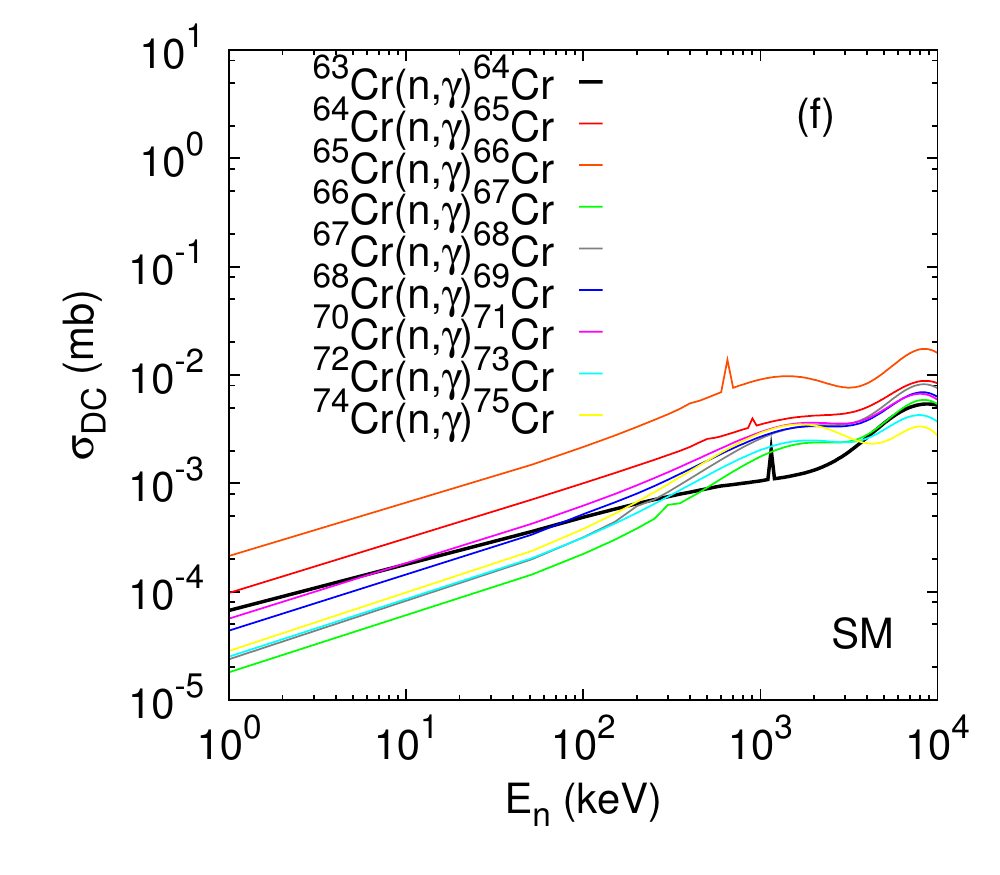}}
\end{figure*}
\begin{figure*}
\resizebox{0.95\textwidth}{!}{
\includegraphics{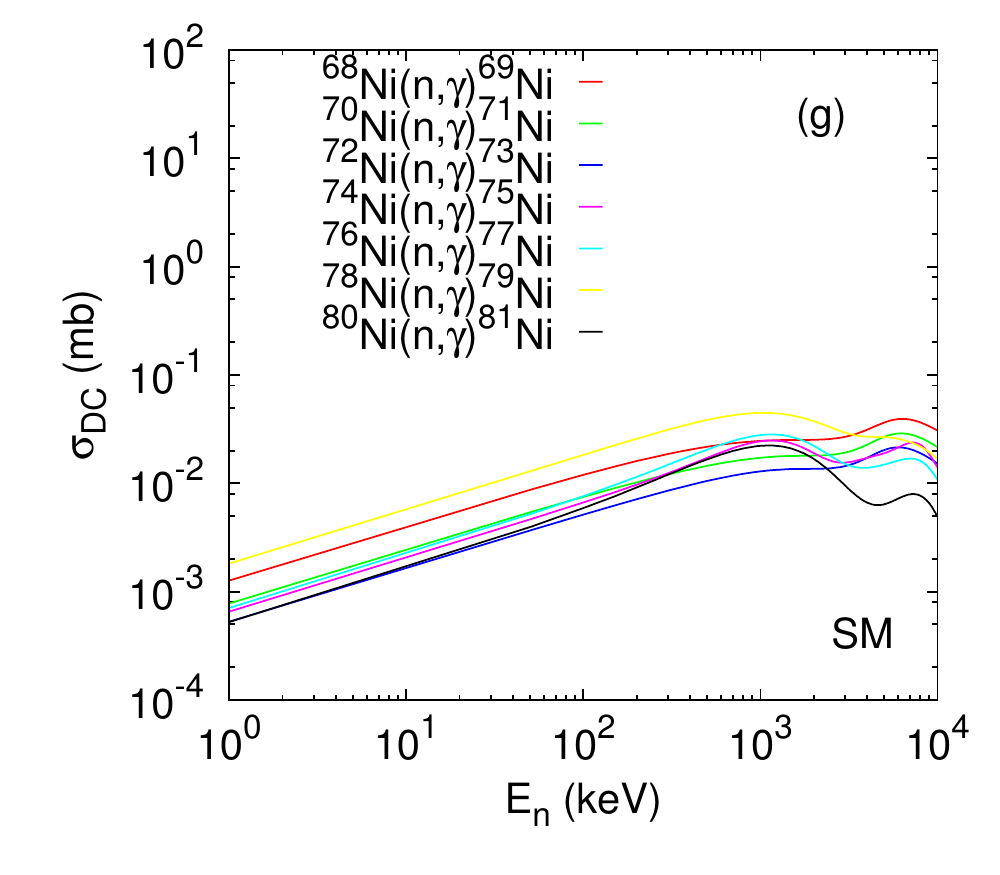}\includegraphics{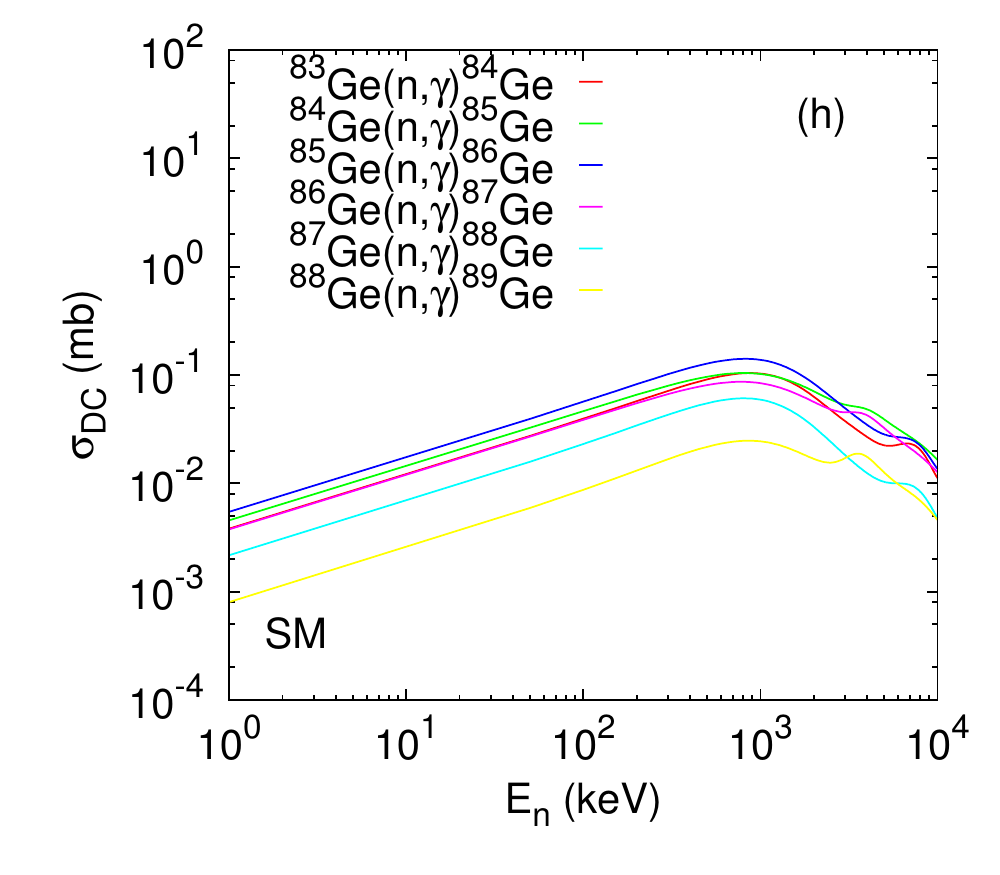}\includegraphics{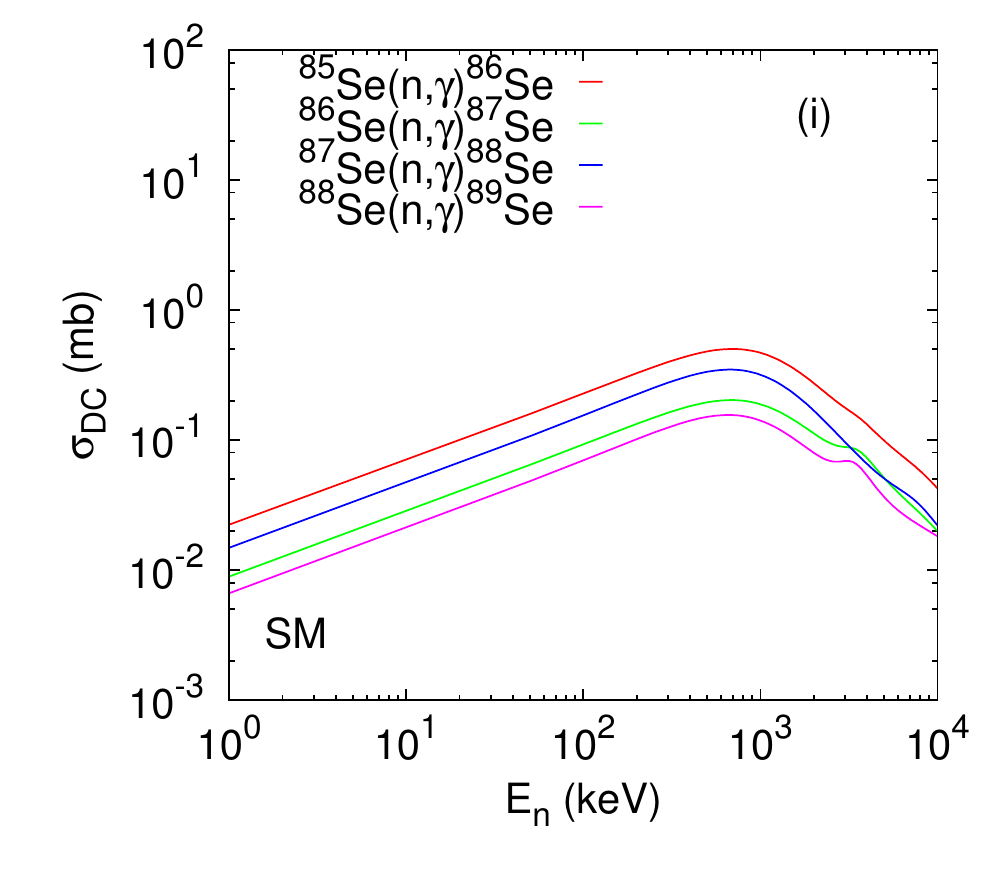}}
\caption{Compilation of theoretical predictions for neutron DC cross sections based on shell-model structure input.
See text for more details.}
\label{dc_SM}
\end{figure*}

The opening of the model space starts playing an important role in nickel nuclei only after 
$N=46$, enhancing the capture cross section by a factor 2 in $^{76}$Ni
up to a factor of 6 in $^{82}$Ni. In deformed Cr nuclei a factor of 2 difference is noted already
at $N=44$ but the discrepancy between different model space calculations 
does not exceed a factor of 3 within the whole Cr isotopic chain, due to a larger fragmentation of wave functions
in deformed nuclei.
The increase of the cross section in both chains is caused by the presence of the $s_{1/2}$
orbital with a significant spectroscopic factor (from 0.2 in $^{73}$Ni to 0.9 in $^{79}$Ni) 
within the $Q$-window. The next positive parity
state, $g_{7/2}$, is located too high to have large fragments in the $Q$-window in the lighter nickel isotopes 
and is also predicted outside the window past the $N=50$ shell closure. It is only in $^{79}$Ni that a $7/2^+$ calculated at 1.22~MeV
with a spectroscopic factor of 0.87 enters the cross section calculation which explains its relatively large value. 
In all cases the trend of the DC cross section with neutron energy remains
unchanged in the astrophysically relevant energy range, and is dominated by the capture on the $d_{5/2}$
orbital, as shown in Fig. \ref{nicr-dc} for the Ni and Cr $N=50$ isotones.   
In the discussion we adopt the results from the LNPS-GDS calculations for $N\ge44$
and keep the results in the LNPS model space for lighter Ni and Cr nuclei.

In Fig.~\ref{dc_SM}, we collect the DC predictions based on shell-model calculations
for all considered nuclei. For transparency, the $sd-pf$ nuclei (panels (a)-(d)) are shown separately for 
odd and even targets. The capture cross section on an odd target with $N$ neutrons
is generally larger than that on the even target with $N+1$ neutrons in all nuclei, varying by a factor 1.2 to 3.5 from case to case and with the neutron incident energy.   
No shell effect is present in the cross section at $N=28$ which is compatible with the 
weakening of the $N=28$ shell closure predicted by the SDPF-U interaction from the shape coexistence in $^{44}$S,
see e.g. Ref.~\cite{Force-S44}, or from the results of Ref.~\cite{Gaudefroy2006} 
in $^{46}$Ar based on the $^{46}$Ar$(d,p)^{47}$Ar transfer reaction.   
On the contrary, one notes an order of magnitude drop in the cross section 
for $^{50,51}$Ar and $^{46}$S targets which is related to the $N=32$ sub-shell closure.
The dependence of the cross section on the incident neutron energy is the same in all isotopes 
and compatible with the $l=1$ neutrons as the shell model predicts large $p_{3/2}$ and $p_{1/2}$
fragments at the lowest excitation energies.
The same dependence is observed in the lightest Ni and Cr isotopes
shown in panel (e) of Fig.~\ref{dc_SM}. Below the $N=40$ the capture still takes place predominantly
on the negative parity states though the $9/2^+$ level, carrying a substantial spectroscopic
factor, appears at low energy. 
For $N>40$ the cross section drops again by several orders of magnitude and changes the tendency
with the incident neutron energy as the capture on the $g_{9/2},d_{5/2}$ orbitals dominates.
As shown previously in Fig. \ref{LNPSvsLNPS}, the capture on the $s_{1/2}$ orbital adds to the cross section but 
does not alternate the neutron-energy dependence. 
The DC cross sections are larger by one order of magnitude in spherical Ni than in deformed Cr isotopes.     
One can also note the drops between the rates on $^{68}$Ni and $^{70}$Ni and $^{78}$Ni and $^{80}$Ni
targets, related to the $N=40$ and $N=50$ shell closures, respectively. 
In panels (h) and (i) we show deformed Ge and Se nuclei having neutron numbers between $N=50$ and $N=56$,
thus with valence particles occupying predominantly the $d_{5/2}$ neutron shell. 
Here, all the positive parity orbitals of the valence space
($d_{5/2,3/2}$, $g_{7/2}$, $s_{1/2}$) with significant spectroscopic factors are predicted  
within the $Q$-window in all considered isotopes. A next drop in the cross section can be noted 
for the $^{88}$Ge related to the $N=56$ closure. A similar calculation in $^{90}$Se could not be performed
due to the computing complexity. 
   
\begin{figure}
\resizebox{0.45\textwidth}{!}{
\includegraphics{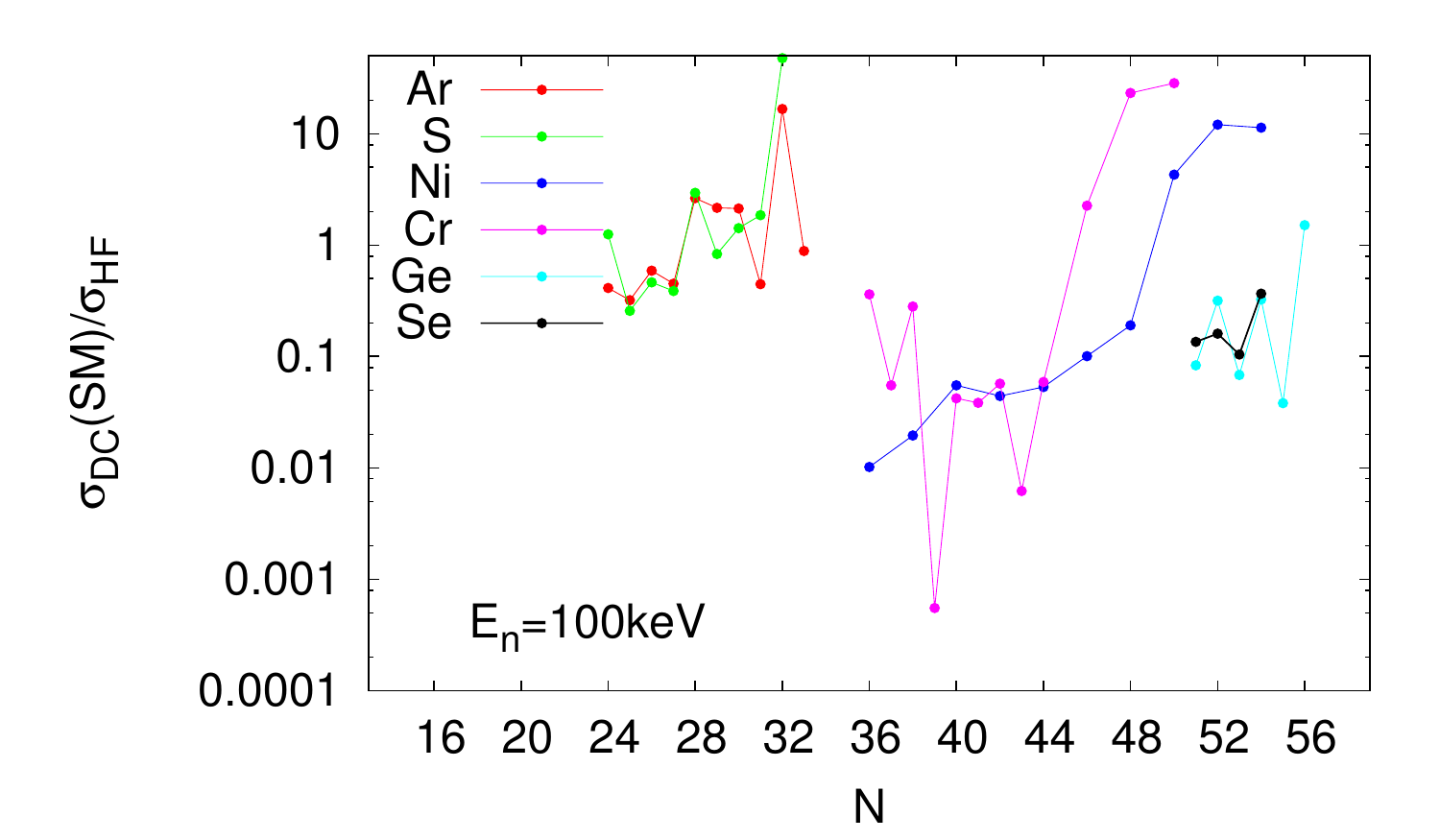}}
\caption{Ratio of direct neutron capture cross sections calculated with the shell-model structure input (SM) to 
the HF calculation at the neutron incident energy of 100keV. Both calculations are obtained
with the TALYS code \cite{TALYS} with the same input, as described in the text.}
\label{ratio}
\end{figure} 

Finally, we compare our SM-based DC cross sections to the resonant 
cross sections which are usually calculated within the statistical HF approach.
The HF model can be questioned near shell closures and in the most neutron-rich region
where the condition of high 
NLD is not maintained and the predicted value can be off by several orders of magnitude.
An alternative description of the resonant neutron capture could be provided on the basis of the High Fidelity Resonance method
treating the resolved resonance region in a statistical approximation \cite{Rochman17} or 
by the shell model, as done e.g. in Ref.~\cite{Fisker2001} for proton-rich nuclei. As in DC, such calculations
require the knowledge of energy levels, spectroscopic factors and $\Gamma$ widths. Nonetheless, while 
electromagnetic transitions can be obtained within the potential model for the DC, 
such a strategy is not desirable for transitions between resonances and bound states, which will
depend strongly on nuclear structure effects. The calculations of transition strengths in the shell model
become complex for the parity-changing $E1$ operator, which requires
enlarged configuration spaces with respect to those used here and go thus beyond the scope of the present
work. Thus, we estimate here the DC and HF cross sections with  
the TALYS code \cite{TALYS} both on the same footing, {\it i.e.} using the same nuclear 
input and in particular the same optical potential (see Ref.~\cite{Xu14} for more details).
Both cross sections are compared in Fig. \ref{ratio}. As expected, the importance of the DC grows with the neutron excess and is predicted to dominate
over the HF part for most neutron-rich nuclei, 
notably the ratio $\sigma_{DC}/\sigma_{HF}$ reaches a factor of 10 after passing the $N=50$
shell closure in Ni isotopes. 

\section{Global microscopic models versus shell-model predictions}
\label{sec:global}
In the following we compare the DC cross sections based on the shell-model structure input 
to the results obtained with the theoretical models described in Ref. \cite{Xu2012}. The approaches of Ref. \cite{Xu2012}
were tested in nuclei for which experimental data for the capture are available and were shown to agree fairly 
with experiment and among them, providing a larger mean value of the spectroscopic factor is used 
when 1p-1h levels only are taken into account with respect to a full combinatorial model of NLD. 
At the same time, the viability of extrapolating global approaches while going to
exotic nuclei could not be judged due to the missing experimental information. The shell-model predictions of discrete levels
and spectroscopic factors far from stability seem to be the most reliable probe at our disposal today to validate other theoretical models. 

\begin{figure*}
\resizebox{0.95\textwidth}{!}{
\includegraphics{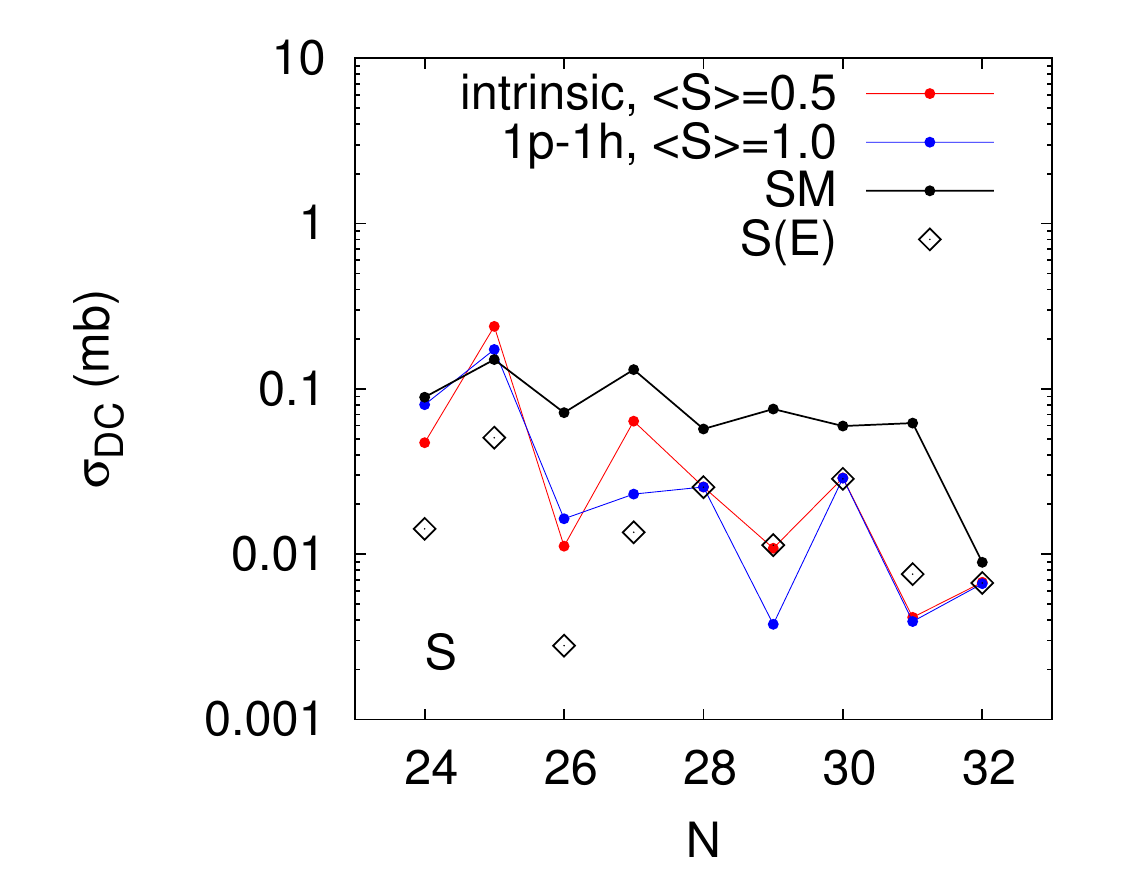}\includegraphics{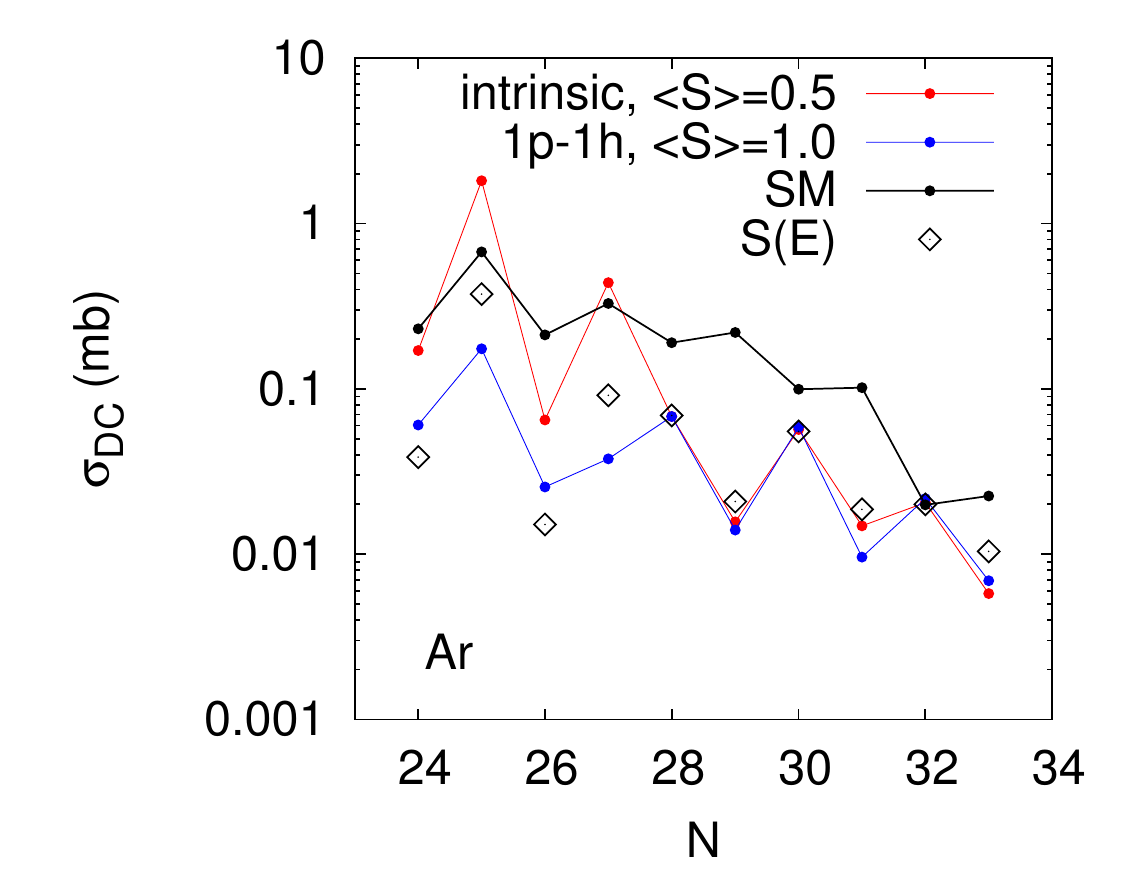}\includegraphics{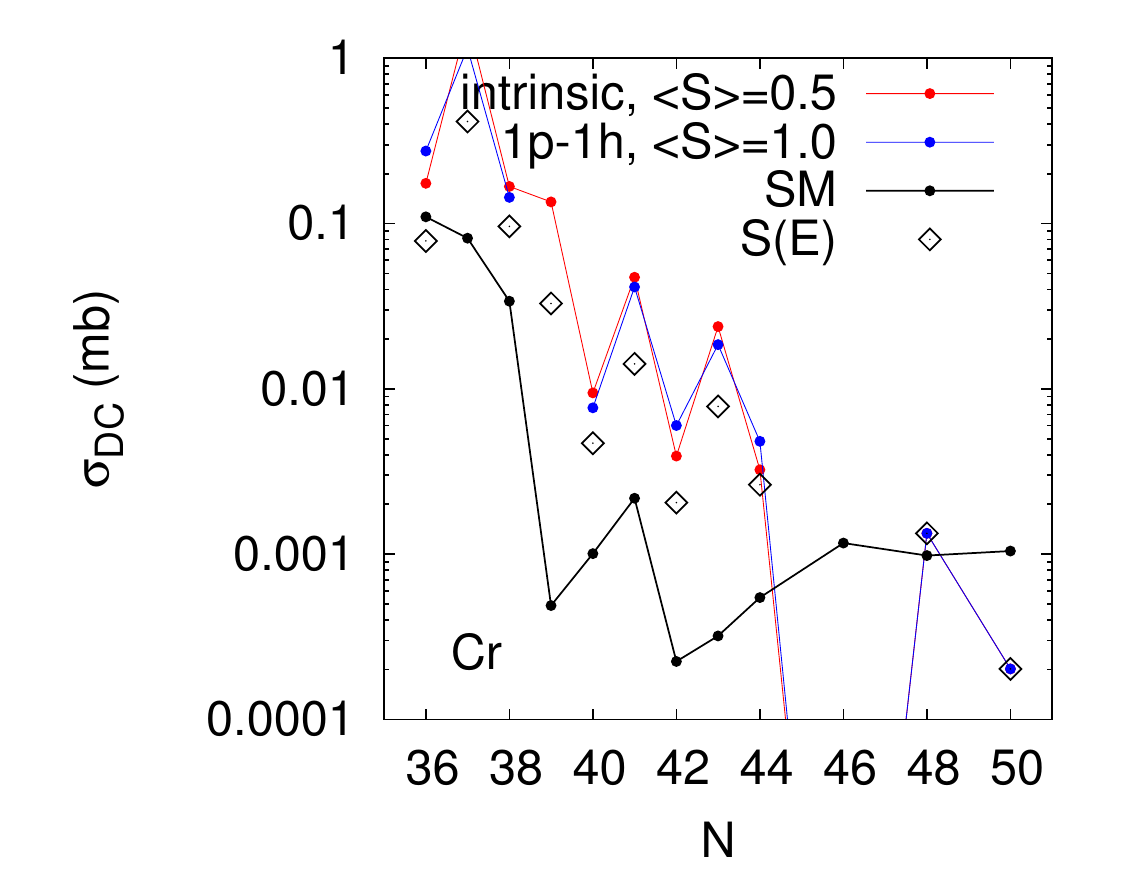}}
\end{figure*}
\begin{figure*}
\resizebox{0.95\textwidth}{!}{
\includegraphics{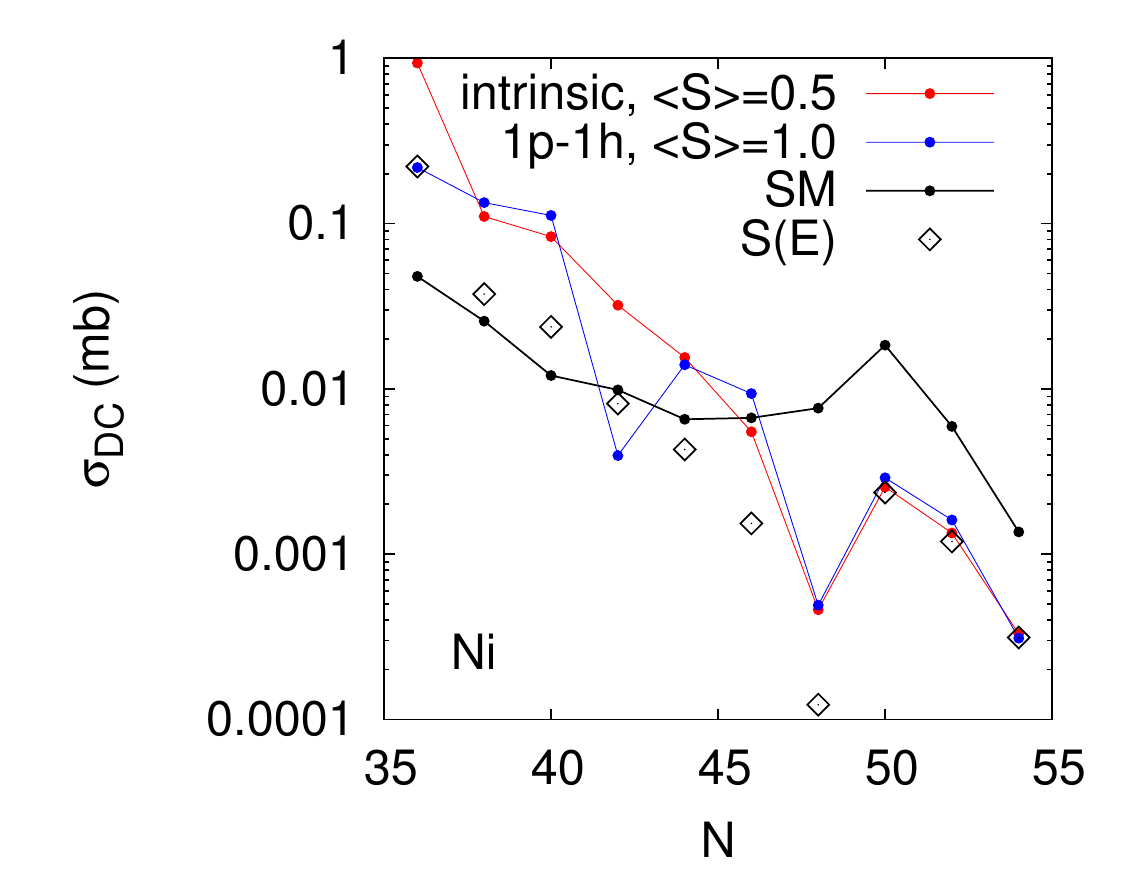}\includegraphics{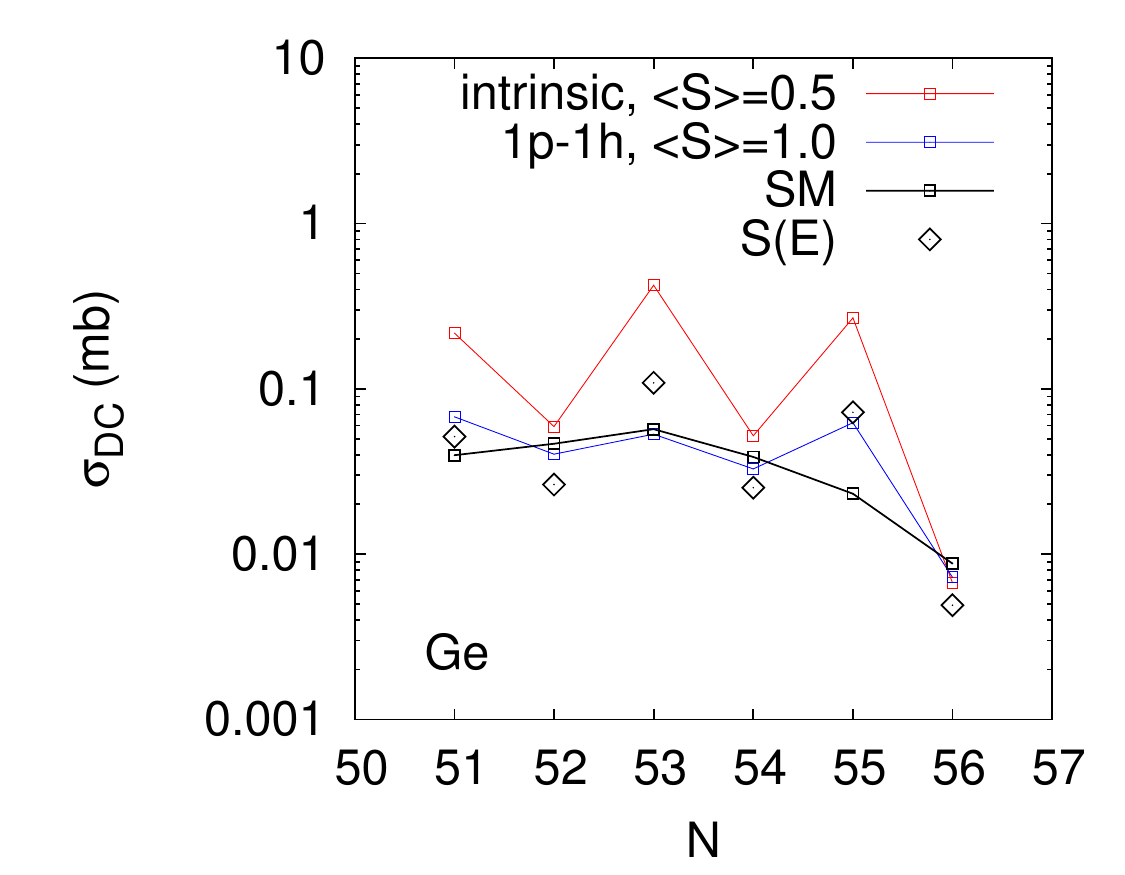}\includegraphics{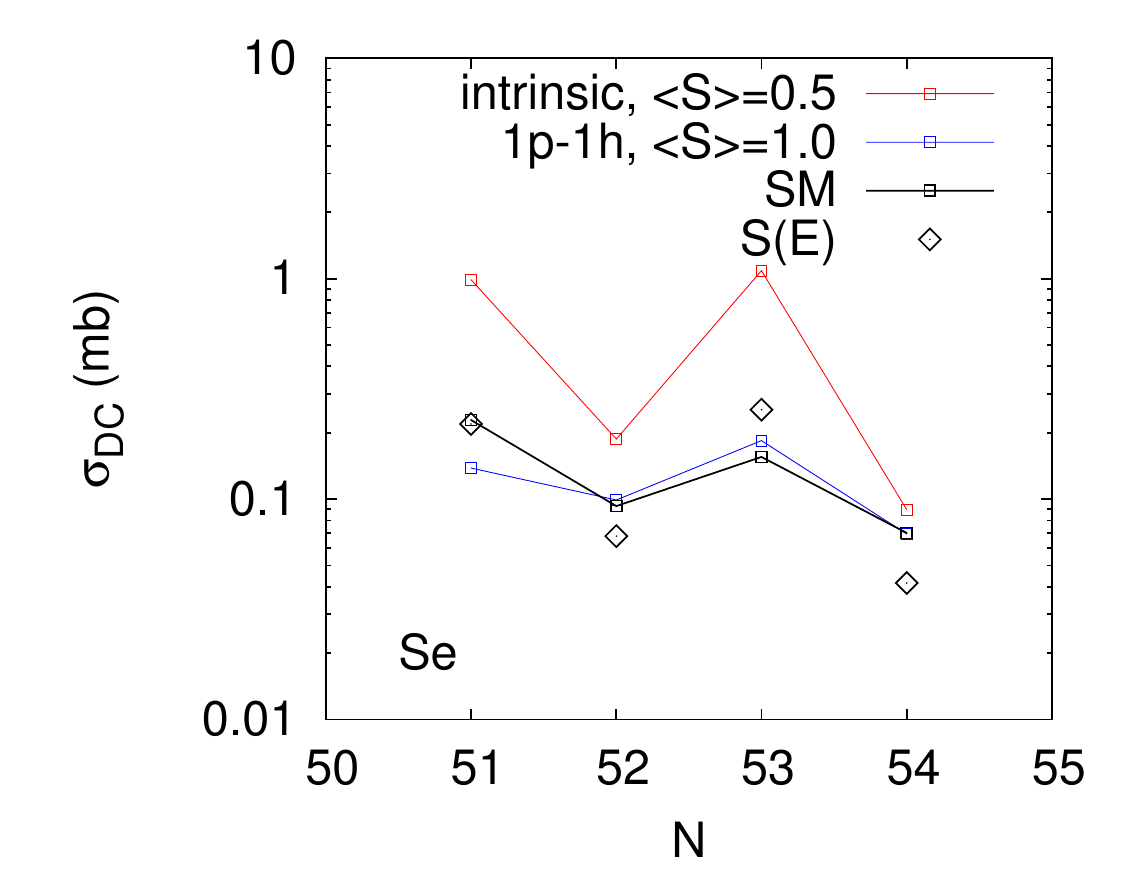}}
\caption{Neutron DC cross sections at the incident neutron energy $E_n=100$~keV calculated with different theoretical
predictions of energy levels and spectroscopic factors for a series of isotopic chains. These include the intrinsic model with $\langle S_f\rangle=0.5$, the 1p-1h model with $\langle S_f\rangle=1$, the SM and the intrinsic model with energy-dependent spectroscopic factor $S(E)$; see text for details.}
\label{dc_all}
\end{figure*}

Neutron capture cross sections from various calculations are shown in Fig. \ref{dc_all} for the neutron incident energy $E_n=100$~keV.
As seen, both the intrinsic and 1p-1h global models provide similar predictions in the majority of cases, 
as observed in Ref. \cite{Xu2012} for nuclei close to the stability.
No result is reported for the 1p-1h model in the case of the capture on $^{63}$Cr as 
the model does not predict transitions from its ground state esimtated to be $1/2^-$ to low energy states
in $^{64}$Cr (at the $N=40$ closure) leading to a null direct capture cross section. 
Otherwise, the largest disagreement between the global models and the shell model concerns 
the neutron capture on $^{70}$Cr where the cross section
from the global models drops to a value four orders of magnitude smaller than SM, pointing
the sensitivity of the calculated DC cross section to the determination of few available states by nuclear models. 
The HFB model predicts a $7/2^+$ ground state in $^{71}$Cr and the first excited state 
$5/2^+$ lies at the energy of 1.2~MeV. In the shell model, the ground state
is $9/2^+$ with a substantial spectroscopic factor and two large fragments of the $d_{5/2}$
orbital are also located within the $Q$-window. 
Apart from these two cases, the models agree however within one order 
of magnitude at $E_n=100$~keV. The inter-model root mean square deviation defined as
\begin{equation}
f_{rms}= \exp \left[\frac{1}{n} \sum_{i=1}^{n} \ln^2 \frac{%
\sigma^i_{SM}}{\sigma^i_{{\rm intrinsic(1p1h)}}}\right]^{1/2} 
\label{frms}
\end{equation}
gives for our 50 nuclei at $E_n=100$~keV a value of 12.26 for the intrinsic model and 11.34 for 1p-1h model, respectively.
The observed discrepancies do not exhibit any particular trend with mass or neutron number,
the shell-model-based cross sections are larger in light nuclei but not necessarily in the mid-mass region.
The agreement between the models is not worse at the neutron-rich side neither.
The only visible tendency is a larger staggering between cross sections on even-even and odd-even targets from the global models
as compared to smoother trends from the SM calculations. Some of that staggering
can be traced back to the adopted 
value of $\langle S_f\rangle$, the same for odd and even targets and for all levels in the $Q$-window
which is not the case in shell-model calculations, as discussed below. 
In Ge and Se, the 1p-1h model happens to follow closely the SM predictions contrary to the 
intrinsic model which overshoots the cross sections for even-odd targets. These nuclei require however more attention.
While 1p-1h model seems to agree well with the shell-model at $E_n=100$~keV,
(Fig. \ref{dc_all}), a closer inspection of these nuclei shows the opposite energy dependence of the calculated 
DC values. SM predicts the increase of the 
cross section with the incident neutron energy, as expected given the availability of the   
$d_{5/2,3/2}$ and $g_{7/2}$ orbitals for the neutron capture. Both global models 
give a bit larger cross sections for the lowest neutron energies and decrease slightly 
with increasing incident neutron energies. As mentioned, the SM predicts non-axial degrees of freedom
being important in $N=52-54$ Ge and Se isotopes, which is supported by the beyond 
mean-field calculations \cite{sieja2013} and experimental evidence \cite{Litzinger,Lettman,czerwinski}.
The present NLD calculations are based on axially-deformed HFB approach
which predicts spherical shapes in $^{84}$Ge and $^{86}$Se. The $2^+$ energies in the intrinsic model
are around $1.6$MeV in both nuclei against $\sim$0.8MeV calculated in SM 
(experimental values being 624~keV and 704~keV, respectively). The agreement between spectra from the intrinsic
model and SM gets better for higher
$N$ values though SM does not predict any spectroscopic factor   
larger than 0.3 in the $Q$-energy window of even-even residual nuclei which explains 
the overestimate by the intrinsic model. Unfortunately, there is no NLD model
based on non-axially deformed mean-field calculations available which could be used in the present cross section calculations.

Apart from differences related to divergent predictions of the lowest-energy levels by theoretical approaches,
the inter-model differences can be further explained by a closer inspection
of the summed spectroscopic strength per energy interval from the shell-model calculations. 
First, let us note that the average spectroscopic factor
from the shell-model calculations for ground states of all even-odd compound nuclei calculated here is 0.364, a value 
very close to the value of 0.347 assumed in the global models from the compilation of all the experimentally 
known spectroscopic factors \cite{Goriely98}.
However, a larger value is found for the even-even residual nuclei, leading to an overall mean value of $S_f\sim0.6$. 
Second, the averaged spectroscopic strength from the SM is not constant with excitation energy: apart the fluctuations related
to the detailed nuclear structure, the model does not predict much strength above a 5~MeV excitation energy
and the amount of spectroscopic strength decreases exponentially. This is a consequence
of using one major shell valence space for neutrons, a choice justified however by 
the existence of well-preserved shell gaps. In the case of the LNPS-GDS model space, 
where the low-energy neutrons can be captured on two major shells due to the presence of intruders, the 
spectroscopic strength is found to decrease with the excitation energy in a similar manner.

\begin{figure}
\resizebox{0.45\textwidth}{!}{\includegraphics{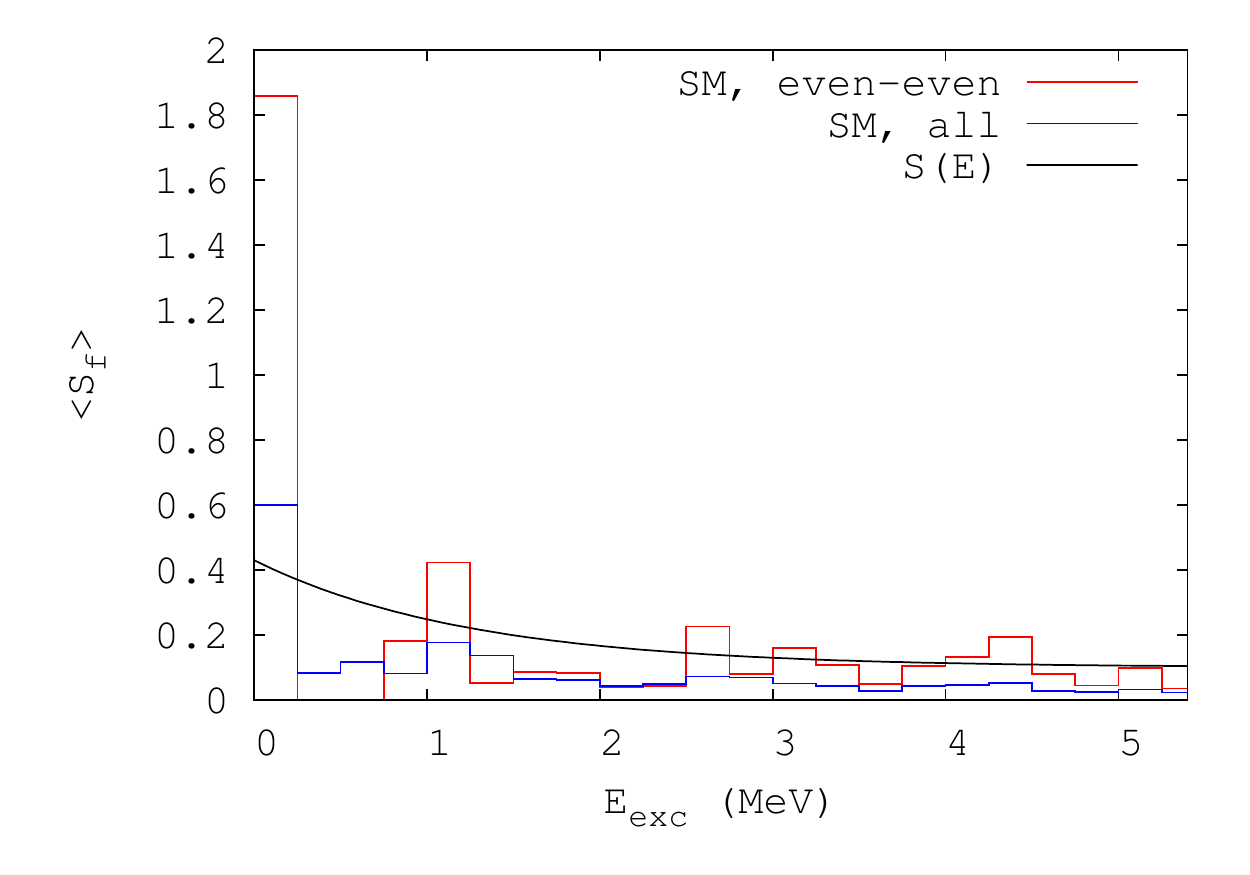}}
\caption{Spectroscopic strength from the shell model calculations per 0.25~MeV 
excitation energy interval, averaged for all studied nuclei (SM, all) or for even-even residual nuclei
only (SM, even-even). The black solid line shows the exponential prescription $S(E)=0.1+0.33 \exp(-0.8E)$ to be used with the intrinsic model of NLD, see
text for details.}
\label{fig-SM}
\end{figure}

In Fig.~\ref{fig-SM} we show the spectroscopic strength summed per energy interval of 0.25~MeV
and averaged for all studied nuclei, or for even-even residual nuclei only. 
It is obvious from that figure that ascribing the same value of $\langle S_f\rangle$ to all excitation energies
may not be realistic and, as noted from Fig.~\ref{dc_all}, it leads to too large cross sections especially in deformed nuclei
with $A\ge 60$. A simple energy-dependent prescription $S(E)=0.1+0.33 \exp(-0.8E)$ (where $E$ is the excitation energy in MeV) based on SM results can be used to improve the predictions in neutron-rich nuclei, as shown in Fig.~\ref{fig-SM}. 
For the ground states, we kept the initial prescription $S_f=0.347$ for even-even targets but we enhanced it to 1.0 for even-odd ones. 
Combined with the intrinsic model of NLD, it permits to reduce 
the inter-model deviation from 12.26 to 9.23. 
The results of the DC cross section calculation based on this prescription are plotted in Fig. \ref{dc_all} and labelled $S(E)$. 
One can see that while the overall improvement
is significant, using $S(E)$ gives a slightly better or equivalent description passed the $N=28$ isotopes in light nuclei
and the improvement is largest in mid-mass deformed nuclei, as the magnitude 
and the staggering of the cross sections are shrunk. In 
light nuclei closer to stability using the $\langle S_f\rangle=0.5$ values gives on average a better fit to SM results.

\section{Application to the r-process nucleosynthesis}

\begin{figure}
\resizebox{0.45\textwidth}{!}{\includegraphics{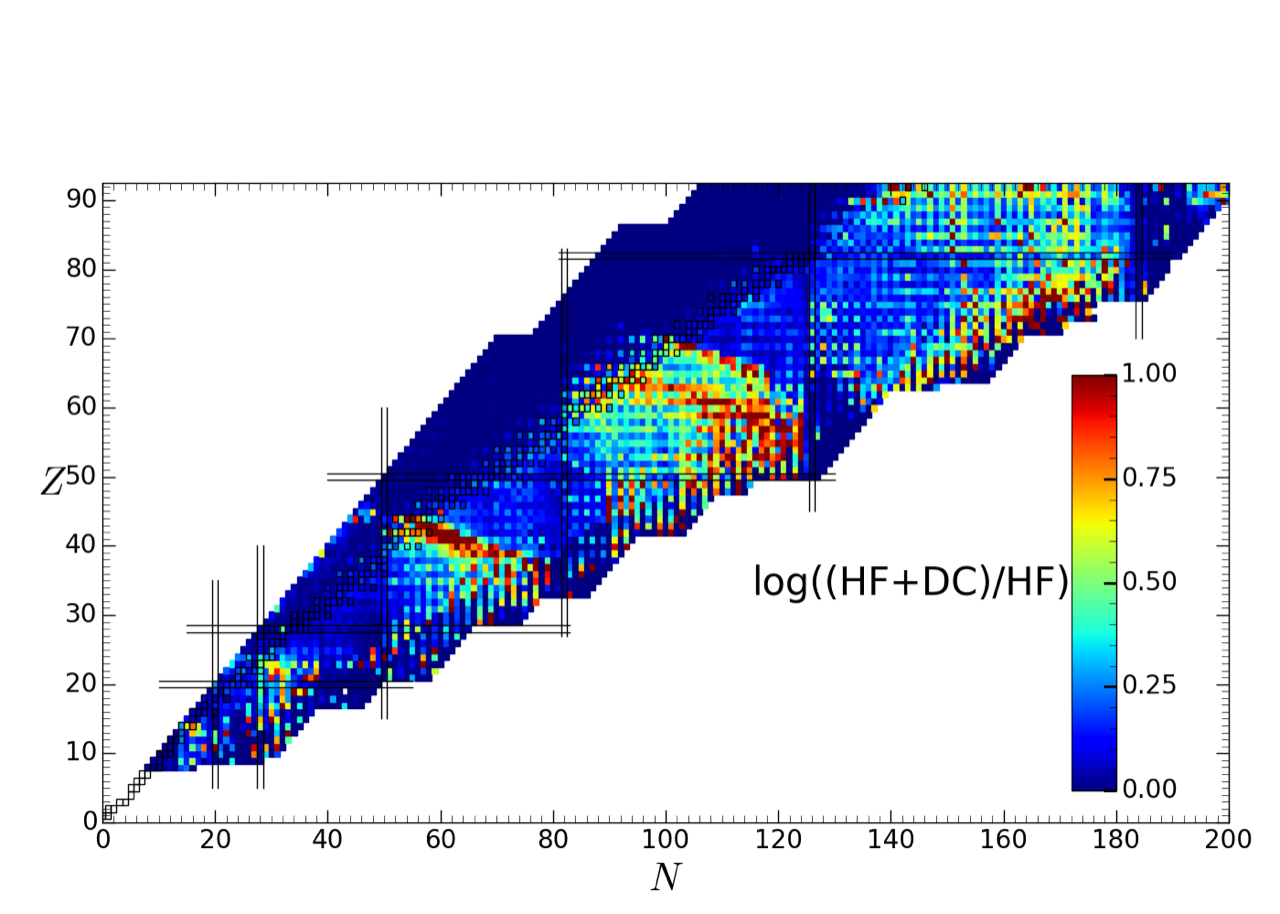}}
\caption{Color-coded representation in the $(N,Z)$ plane of the ratio (in log) between the $(n,\gamma)$ reaction rate at $T=10^9$~K obtained with the HF plus intrinsic DC using an energy-dependent spectroscopic factor $S(E)$ to the one neglecting the DC contribution. Ratios above 10 are shown in red. All nuclei with $8 \le Z \le 92$ lying between the HFB-21 proton and neutron drip lines  are represented.}
\label{fig_ng_int_NZ}
\end{figure}

\begin{figure}
\resizebox{0.45\textwidth}{!}{\includegraphics{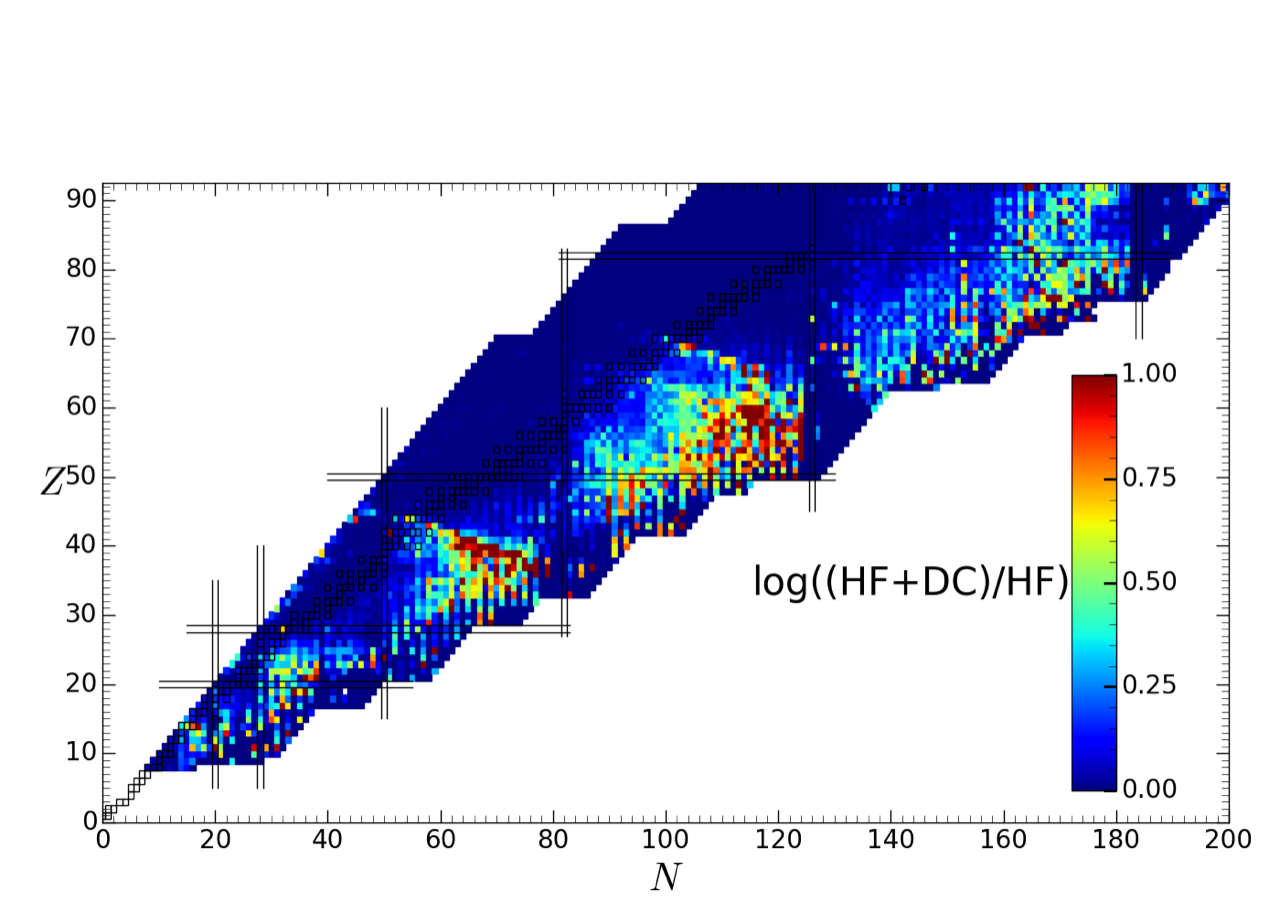}}
\caption{Same as Fig.~\ref{fig_ng_int_NZ} when the DC is calculated with the 1p-1h model and a constant spectroscopic factor $S_f=1$.}
\label{fig_ng_1p1h_NZ}
\end{figure}

The  neutron capture rates of astrophysics interest have been calculated for about 5800 nuclei with $8 \le Z \le 110$ lying between the proton and neutron drip lines. Figs.~\ref{fig_ng_int_NZ}-\ref{fig_ng_1p1h_NZ} illustrate the ratio between the total (HF+DC) and the HF reaction rates in the $(N,Z)$ plane, when the DC is either calculated by the intrinsic model with a energy-dependent spectroscopic factor or by the 1p-1h model with a constant $S_f=1$ (as described in Sec.~\ref{sec:global}). In both cases, the DC contribution increases the radiative neutron capture rate by a factor up to 100 for drip line nuclei. As already pointed out in Refs.~\cite{Xu2012,Xu14,Goriely1997},  for some neutron-rich nuclei, no allowed direct transitions may be found (due to selection rules), and the direct channel can consequently be inhibited. However,  neutron-rich nuclei with $N\lsimeq 82$ or 126, radiative neutron capture rates including DC contributions are seen to be significantly larger with respect to the HF-only prediction.

As an illustration of the impact of the newly derived radiative neutron capture rates, r-process nucleosynthesis calculations have been performed. The impact of reaction rates on the r-process nucleosynthesis remains difficult to ascertain in the sense that their influence strongly depends on the adopted astrophysical scenario and most particularly on the temperature at which the r-process takes place \cite{Arnould07}. In particular, at low temperatures (typically below $10^9$~K), photodisintegration rates are slow and consequently no (n,$\gamma$)-($\gamma$,n) equilibrium can be reached. In this case, the neutron capture rates directly influence the calculated abundances \cite{Arnould07}. It should however be recalled here that, so far, all  r-process calculations have made use of neutron capture rates evaluated within the statistical HF model and that the DC component is never included in such calculations.

To test the impact of the newly derived reaction rates including the DC component on the r-process nucleosynthesis, 
simulations of neutron star (NS) mergers have been considered and the composition of the ejecta estimated. 
Both the dynamical  ejecta from the merger and the post-merger outflow from the viscously driven wind 
of the BH -- torus system are considered, as detailed in Ref.~\cite{Goriely11,Goriely15,Just15}.
The dynamical ejecta is calculated for a symmetric 1.365--1.365$M_{\odot}$ 
binary system compatible with the total mass detected in the GW170817 event \cite{Abbott17}.  A total of $4.9\times 10^{-3}~M_{\odot}$ is found to be expelled.
In the nucleosynthesis simulation,  the weak interactions on free nucleons are taken into account in the  parametric approach described in Ref.~\cite{Goriely15} in terms of prescribed neutrino luminosities and mean energies, Values of $L_{\nu_e} =0.3\times 10^{53}$~erg/s; $L_{{\bar\nu}_e}= 10^{53}$~erg/s; $\langle E_{\nu_e}\rangle=8$~MeV; $ \langle E_{{\bar\nu}_e}\rangle = 12$~MeV guided by hydrodynamical simulations including a detailed account of neutrino absorption \cite{Ardevol19} are adopted here. 

In addition, we also estimate here the composition of the material ejected from a system 
characterized by a torus mass of 0.1~$M_{\odot}$ and a 3\,$M_\odot$ BH (corresponding to the M3A8m1a5 model of Ref.~\cite{Just15}). 
The total mass ejected from the BH-torus system amounts to $2.5 \times 10^{-2} M_{\odot}$, 
and the outflow is characterised by a mean electron fraction $\bar{Y_e}=0.24$, a mean entropy 
${\bar s}/k_B=28$ and a mean velocity ${\bar v}=1.56 \times 10^9$~cm/s.
More details about the additional nuclear inputs and the astrophysical scenario can be found in Refs.~\cite{Just15,Lemaitre20}.

\begin{figure}
\resizebox{0.5\textwidth}{!}{\includegraphics{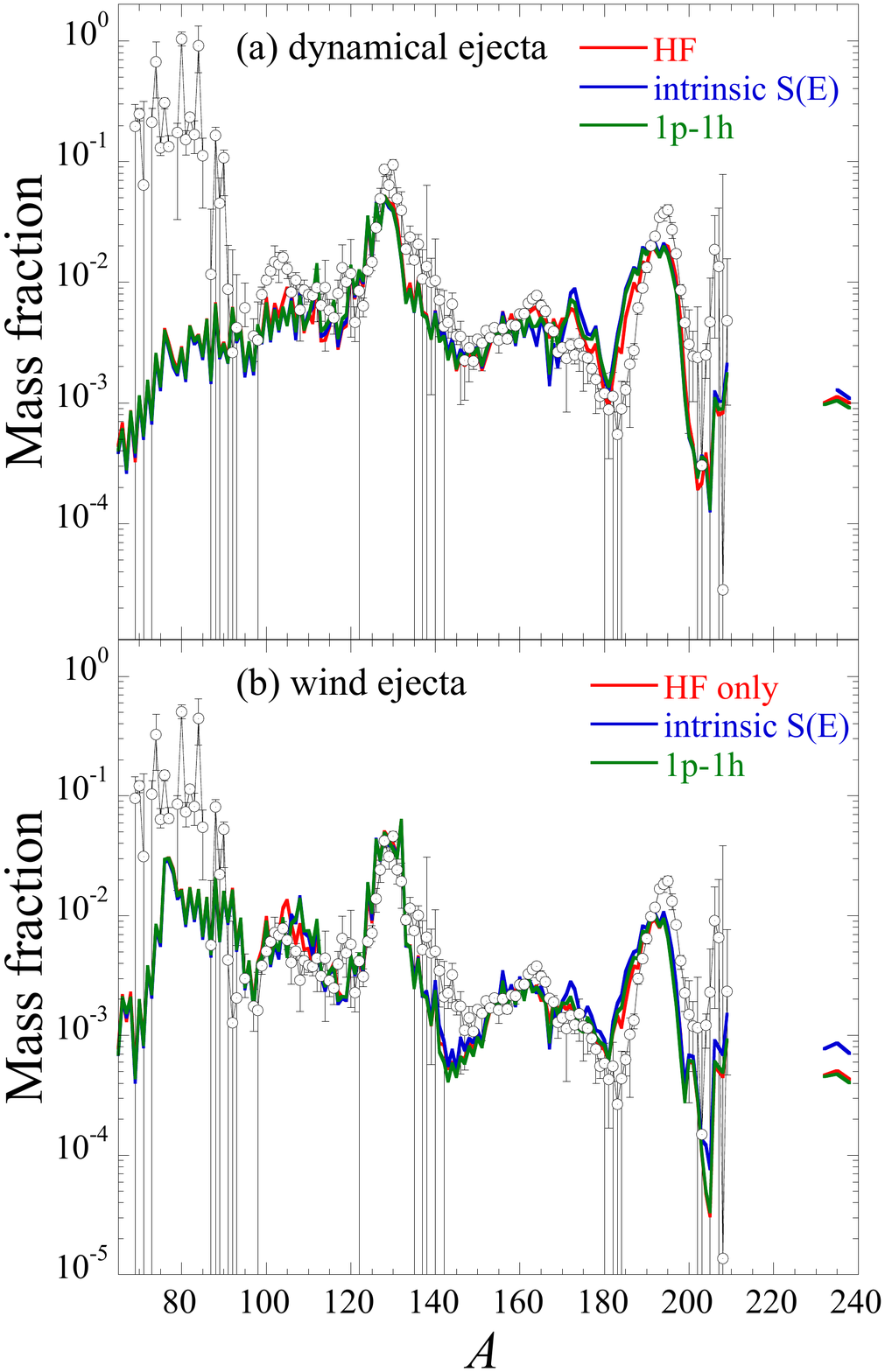}}
\caption{(a) (Color online) Mass fraction of the $4.9\times 10^{-3}~M_{\odot}$ of material dynamically ejected in
         the 1.365--1.365$M_{\odot}$ NS-NS merger model as a function of the 
         atomic mass $A$. The red, blue and green curves are obtained with neutron capture and photoneutron rates derived from HF formalism only, HF plus   intrinsic DC (using an energy-dependent spectroscopic factor $S(E)$) or  HF plus DC calculated with the 1p-1h model and a constant spectroscopic factor $S_f=1$. The arbitrarily normalized solar system r-abundance distribution 
         (open circles) is shown for comparison \cite{Goriely99}.  (b) Same as (a) for the wind ejecta from a   3\,$M_\odot$ BH -- 0.1\,$M_\odot$
 torus system.
}
\label{fig_rpro}
\end{figure}

The final composition of the material ejected in both scenarios is shown in Fig.~\ref{fig_rpro}. In both cases, the 3 sets of reaction rates lead to relatively similar predictions. The inclusion of the DC contribution is seen to affect essentially the low-mass tail of the third r-process peak . Such deviations can be associated with the larger neutron captures rates found for the $N \lsimeq 126$ neutron-rich nuclei, as seen in Figs.~\ref{fig_ng_int_NZ}- \ref{fig_ng_1p1h_NZ}. Both the intrinsic or the 1p-1h prescriptions give rather similar r-abundance distributions, except for the production of actinides in the wind ejecta.
It should be recalled here however that larger deviations can be expected if the HF contribution is not as strong as estimated here. The HF mechanism assumes the number of resonances in the compound system is large enough to ensure an average statistical continuum superposition of available resonances. However, 
for exotic neutron-rich nuclei with a low neutron separation energy, the resonance capture region should be resolved, so that the neutron capture should rather be described within a Breit-Wigner approach or, for example, on the basis of the High Fidelity Resonance method \cite{Rochman17}. In this case, the DC contribution could have a significantly stronger impact on the neutron capture rates than the one found here by treating the resonance capture within the HF model.

\section{Conclusions}
The present study has been devoted to the direct neutron-capture process. The shell-model approach has been used with well-established effective interactions  to estimate the properties of the excitation spectrum and the corresponding spectroscopic factors for a set of 50 neutron-rich target nuclei in different mass regions, including doubly-, semi-magic and deformed ones. 
Those shell-model  energies and spectroscopic 
factors have bee included in TALYS reaction code to evaluate the direct contribution to the neutron capture rates
and to test global theoretical models using average spectroscopic factors and level densities based on the Hartree-Fock-Bogoliubov plus combinatorial method. 

The comparison between shell-model and global model results reveals several discrepancies 
that can be related to problems in level densities and the difficulty to estimate at low energies 1p-1h excitations as well as the total level density.
In particular, the global models of NLD with an average value of spectroscopic factor perform reasonably well when compared to the shell-model predictions, the discrepancies between both models remaining rather similar further from the stability.
The structure effects are important and alternate inter-model differences substantially along the isotopic chains.
It is however possible to
bring models into a closer agreement for the neutron-rich nuclei, by a simple modification of the value of the spectroscopic factor used
with combinatorial model of NLD or by including an energy-dependent spectroscopic factor.
We also note major deviations for nuclei predicted to have $\gamma$-bands, ascribing this problem to the use
of axially-deformed mean field for the construction of the NLD.    
The improvements of the global models for computations of the astrophysical interest 
should thus focus on the accurate description of the mean-field within a larger variational basis 
but also on a proper construction of NLD for non-axial deformations. 

Both global DC models agree that, for exotic neutron-rich nuclei, the direct capture is non-negligible with respect to the HF
contribution and can be up to a factor 100 larger than the resonant capture for drip line nuclei.  Such deviations are however not enough to 
significantly affect the r-abundance distributions predicted to be ejected from binary NS mergers. 
The relevance of the HF description of the resonant capture for neutron-rich nuclei remains unclear 
but can be assessed using shell-model predictions of necessary nuclear structure ingredients. Such a work is 
currently in progress for selected nuclei.

\section{Acknowledgments}
S.G. acknowledges financial support from FNRS (Belgium). This work was partially supported by the Fonds de la Recherche Scientifique - FNRS \
and the Fonds Wetenschappelijk Onderzoek - Vlaanderen (FWO) under the EOS Project No O022818F.

%\section{References}
\bibliographystyle{spphys}
%\bibliography{../../BIB/kama}

\begin{thebibliography}{999}
\providecommand{\url}[1]{{#1}}
\providecommand{\urlprefix}{URL }
\expandafter\ifx\csname urlstyle\endcsname\relax
  \providecommand{\doi}[1]{DOI \discretionary{}{}{}#1}\else
  \providecommand{\doi}{DOI \discretionary{}{}{}\begingroup
  \urlstyle{rm}\Url}\fi
\bibitem{Arnould07}
M. Arnould, S. Goriely, K. Takahashi, Phys. Rep. \textbf{450}, 97 (2007)

\bibitem{Arnould20}
M. Arnould, S. Goriely, Prog. Part. Nucl. Phys. \textbf{112}, 103766 (2020)

\bibitem{Goriely08}
S. Goriely, S. Hilaire, A. J. Koning, Astron. Astrophys. \textbf{487}, 767 (2008)

\bibitem{Satchler80}
G.R. Satchler, in {\it Introduction to nuclear reactions}, Macmillan press ltd. (1980)

\bibitem{Oberhummer91} H. Oberhummer, and G. Staudt, Direct Reaction Mechanism in Astrophysically Relevant Processes, in: H. Oberhummer (Ed.), Nuclei in the Cosmos, Springer Verlag, Heidelberg, 29 (1991).

\bibitem{Mengoni95} A. Mengoni, T. Otsuka, and M. Ishihara, Phys. Rev. C \textbf{52}, R2334 (1995).

\bibitem{Beer96} H. Beer, C. Coceva, P. V. Sedyshev, Y. P. Popov, H. Herndl, R. Hofinger, P. Mohr, and H. Oberhummer, Phys. Rev. C \textbf{54}, 2014 (1996).

\bibitem{Descouvemont03} P. Descouvemont, Theoretical Models for Nuclear Astrophysics, Nova Science Publishers, New York (2003).

\bibitem{Descouvemont08} P. Descouvemont, J. Phys. G \textbf{35}, 014006 (2008).

\bibitem{Xu2012}
Y.~Xu, S.~Goriely, Phys. Rev. C \textbf{86}, 045801 (2012)

\bibitem{Xu14}
Y.~Xu, S.~Goriely, A.J. Koning and S. Hilaire, Phys. Rev. C \textbf{90}, 024604 (2014)

\bibitem{Goriely1997}
S.~Goriely, Astron. Astrophys. \textbf{325}, 414 (1997)

\bibitem{Goriely-LD}
S.~Goriely, S.~Hilaire, A.J. Koning, Phys. Rev. C \textbf{78}, 064307 (2008)

\bibitem{Goriely98}
S.~Goriely, Phys. Lett. B \textbf{436}, 10 (1998)

\bibitem{KD}
A.~Koning, J.~Delaroche, Nuclear Physics A \textbf{713}(3), 231  (2003)

\bibitem{Goriely-HFB27}
S.~Goriely, N.~Chamel, J.M. Pearson, Phys. Rev. C \textbf{88}, 061302(R) (2013)

\bibitem{ANTOINE}
E.~Caurier, F.~Nowacki, Acta Phys. Pol. \textbf{B30}, 705 (1999)

\bibitem{RMP}
E.~Caurier, G.~Martinez-Pinedo, F.~Nowacki, A.~Poves, A.P. Zuker, Rev. Mod.
  Phys. \textbf{77}, 427 (2005)

\bibitem{SDPF}
F.~Nowacki, A.~Poves, Phys. Rev. C \textbf{79}(1), 014310 (2009)

\bibitem{Bastin2007}
B.~Bastin, et~al., Phys. Rev. Lett. \textbf{99}, 022503 (2007).
\newblock \doi{10.1103/PhysRevLett.99.022503}

\bibitem{Gaudefroy2006}
L.~Gaudefroy, O.~Sorlin, D.~Beaumel, Y.~Blumenfeld, Z.~Dombr\'adi {\it et al.},
%S.~Fortier,
%  S.~Franchoo, .M. G\'elin, J.~Gibelin, S.~Gr\'evy, F.~Hammache, F.~Ibrahim,
%  K.W. Kemper, K.L. Kratz, S.M. Lukyanov, C.. Monrozeau, L.~Nalpas, F.~Nowacki,
%  A.N. Ostrowski, T.~Otsuka, Y.E. Penionzhkevich, J.~Piekarewicz, E.C.
%  Pollacco, P.~\~Roussel-Chomaz, E.~Rich, J.A. Scarpaci, M.G. St.~Laurent,
%  D.~Sohler, M.~Stanoiu, T.~Suzuki, E.. Tryggestad, D.~Verney, 
Phys. Rev. Lett.  \textbf{97}, 092501 (2006)

\bibitem{Gaudefroy2009}
L.~Gaudefroy, J.M. Daugas, M.~Hass, S.~Gr\'evy, C.~Stodel {\it et al.},
% J.C. Thomas,
%  L.~Perrot, M.~Girod, B.~Ross\'e, J.C. Ang\'elique, D.L. Balabanski, E.~Fiori,
%  C.~Force, G.~Georgiev, D.~Kameda, V.~Kumar, R.L. Lozeva, I.~Matea, V.~M\'eot,
%  P.~Morel, B.S.N. Singh, F.~Nowacki, G.~Simpson, 
Phys. Rev. Lett.  \textbf{102}, 092501 (2009).
\newblock \doi{10.1103/PhysRevLett.102.092501}.
\newblock
  \urlprefix\url{https://link.aps.org/doi/10.1103/PhysRevLett.102.092501}

\bibitem{Force-S44}
C.~Force, S.~Gr\'evy, L.~Gaudefroy, O.~Sorlin, L.~C\'aceres {\it et al.},
%, F.~Rotaru,
%  J.~Mrazek, N.L. Achouri, J.C. Ang\'elique, F.~Azaiez, B.~Bastin, R.~Borcea,
%  A.~Buta, J.M. Daugas, Z.~Dlouhy, Z.~Dombr\'adi, F.~De~Oliveira, F.~Negoita,
%  Y.~Penionzhkevich, M.G. Saint-Laurent, D.~Sohler, M.~Stanoiu, I.~Stefan,
%  C.~Stodel, F.~Nowacki, 
Phys. Rev. Lett. \textbf{105}, 102501 (2010)

\bibitem{Santiago-Gonzalez}
D.~Santiago-Gonzalez, I.~Wiedenh\"over, V.~Abramkina, M.L. Avila, T.~Baugher {\it et al.},
%  D.~Bazin, B.A. Brown, P.D. Cottle, A.~Gade, T.~Glasmacher, K.W. Kemper,
%  S.~McDaniel, A.~Rojas, A.~Ratkiewicz, R.~Meharchand, E.C. Simpson, J.A.
%  Tostevin, A.~Volya, D.~Weisshaar, 
Phys. Rev. C \textbf{83}, 061305 (2011)

\bibitem{Liu-Ar52}
H.~Liu, et~al., Phys. Rev. Lett. \textbf{122}(7), 072502 (2019)

\bibitem{Gade-S45}
A.~Gade, et~al., Phys. Rev. C \textbf{93}(5), 054315 (2016)

\bibitem{Calinescu-Ar46}
S.~Calinescu, L.~C\'aceres, S.~Gr\'evy, O.~Sorlin, Z.~Dombr\'adi {\it et al.},
%, M.~Stanoiu,
%R.~Astabatyan, C.~Borcea, R.~Borcea, M.~Bowry, W.~Catford, E.~Cl\'ement,
%S.~Franchoo, R.~Garcia, R.~Gillibert, I.H. Guerin, I.~Kuti, S.~Lukyanov,
%A.~Lepailleur, V.~Maslov, P.~Morfouace, J.~Mrazek, F.~Negoita, M.~Niikura,
%L.~Perrot, Z.~Podoly\'ak, C.~Petrone, Y.~Penionzhkevich, T.~Roger, F.~Rotaru,
%D.~Sohler, I.~Stefan, J.C. Thomas, Z.~Vajta, E.~Wilson, 
Phys. Rev. C  \textbf{93}, 044333 (2016)

\bibitem{Saxena}
P.~Srivastava, J.G. Hirsch, M.~Ermamatov, V.~Kota, Nucl. Phys. A \textbf{961},
  68 (2017)

\bibitem{Chevrier}
R.~Chevrier, et~al., Phys. Rev. Lett. \textbf{108}, 162501 (2012).
\newblock \doi{10.1103/PhysRevLett.108.162501}

\bibitem{Lenzi2010}
S.M. Lenzi, F.~Nowacki, A.~Poves, K.~Sieja, Phys. Rev. C \textbf{82}(5), 054301
  (2010).
\newblock \doi{10.1103/PhysRevC.82.054301}

\bibitem{Ljungvall}
J.~Ljungvall, et~al., Phys. Rev. C \textbf{81}(6), 061301 (2010).
\newblock \doi{10.1103/PhysRevC.81.061301}

\bibitem{Gade-Ti60}
A.~Gade, et~al., Phys. Rev. Lett. \textbf{112}, 112503 (2014)

\bibitem{Modamio}
V.~Modamio, et~al., Phys. Rev. C \textbf{88}, 044326 (2013)

\bibitem{Diriken}
J.~Diriken, et~al., Physics Letters B \textbf{736}(0), 533  (2014)

\bibitem{Georgi2011}
E.~Fiori, et~al., Phys. Rev. C \textbf{85}, 034334 (2012)

\bibitem{Elisa2011}
E.~Rapisarda, et~al., Phys. Rev. C \textbf{84}, 064323 (2011)

\bibitem{Eda}
E.~Sahin, et~al., Phys. Rev. C \textbf{91}, 034302 (2015)

\bibitem{Vajta}
Z.~Vajta, D.~Sohler, Y.~Shiga, K.~Yoneda, K.~Sieja {\it et al.},
%, D.~Steppenbeck,
%  Z.~Dombrodi, N.~Aoi, P.~Doornenbal, J.~Lee, H.~Liu, M.~Matsushita,
%  S.~Takeuchi, H.~Wang, H.~Baba, P.~Bednarczyk, Z.~Fulop, S.~Go,
%  T.~Hashimoto, E.~Ideguchi, K.~Ieki, K.~Kobayashi, Y.~Kondo, R.~Minakata,
%  T.~Motobayashi, D.~Nishimura, H.~Otsu, H.~Sakurai, Y.~Sun, A.~Tamaii,
%  R.~Tanaka, Z.~Tian, T.~Yamamoto, X.~Yang, Z.~Yang, Y.~Ye, R.~Yokoyama,
%  J.~Zenihiro, 
Physics Letters B \textbf{782}, 99  (2018)

\bibitem{Dijon}
A.~Dijon, et~al., Phys. Rev. C \textbf{85}, 031301 (2012)

\bibitem{Co-76}
P.A. S\"oderstr\"om, S.~Nishimura, Z.Y. Xu, K.~Sieja, V.~Werner {\it et al.},
%, P.~Doornenbal,
%  G.~Lorusso, F.~Browne, G.~Gey, H.S. Jung, T.~Sumikama, J.~Taprogge, Z.~Vajta,
%  H.~Watanabe, J.~Wu, H.~Baba, Z.~Dombradi, S.~Franchoo, T.~Isobe, P.R. John,
%  Y.K. Kim, I.~Kojouharov, N.~Kurz, Y.K. Kwon, Z.~Li, I.~Matea, K.~Matsui,
%  G.~Mart\'{\i}nez-Pinedo, D.~Mengoni, P.~Morfouace, D.R. Napoli, M.~Niikura,
%  H.~Nishibata, A.~Odahara, K.~Ogawa, N.~Pietralla,
%  E.~\ifmmode~\mbox{\c{S}}\else \c{S}\fi{}ahin, H.~Sakurai, H.~Schaffner,
%  D.~Sohler, I.G. Stefan, D.~Suzuki, R.~Taniuchi, A.~Yagi, K.~Yoshinaga, 
Phys.  Rev. C \textbf{92}, 051305 (2015)

\bibitem{Meisel}
Z.~Meisel, S.~George, S.~Ahn, D.~Bazin, B.A. Brown {\it et al.},
%, J.~Browne, J.F. Carpino,
%  H.~Chung, R.H. Cyburt, A.~Estrad\'e, M.~Famiano, A.~Gade, C.~Langer,
%  M.~Mato\ifmmode~\check{s}\else \v{s}\fi{}, W.~Mittig, F.~Montes, D.J.
%  Morrissey, J.~Pereira, H.~Schatz, J.~Schatz, M.~Scott, D.~Shapira, K.~Sieja,
%  K.~Smith, J.~Stevens, W.~Tan, O.~Tarasov, S.~Towers, K.~Wimmer, J.R.
%  Winkelbauer, J.~Yurkon, R.G.T. Zegers, 
Phys. Rev. C \textbf{93}, 035805  (2016)

\bibitem{Shand}
C.~Shand, Z.~Podolyak, M.~Gorska, P.~Doornenbal, A.~Obertelli, {\it et al.}
%  F.~Nowacki,
%  T.~Otsuka, K.~Sieja, J.~Tostevin, Y.~Tsunoda, G.~Authelet, H.~Baba,
%  D.~Calvet, A.~Chateau, S.~Chen, A.~Corsi, A.~Delbart, J.~Gheller,
%  A.~Giganon, A.~Gillibert, T.~Isobe, V.~Lapoux, M.~Matsushita, S.~Momiyama,
%  T.~Motobayashi, M.~Niikura, H.~Otsu, N.~Paul, C.~Piron, A.~Peyaud,
%  E.~Pollacco, J.Y. Rousso, H.~Sakurai, C.~Santamaria, M.~Sasano, Y.~Shiga,
%  D.~Steppenbeck, S.~Takeuchi, R.~Taniuchi, T.~Uesaka, H.~Wang, K.~Yoneda,
%  T.~Ando, T.~Arici, A.~Blazhev, F.~Browne, A.~Bruce, R.~Carroll, L.~Chung,
%  M.~Cortº©s, M.~Dewald, B.~Ding, Z.~Dombrodi, F.~Flavigny, S.~Franchoo,
%  F.~Giacoppo, A.~Gottardo, K.~HadyèÑska-KlÒôk, A.~Jungclaus, Z.~Korkulu,
%  S.~Koyama, Y.~Kubota, J.~Lee, M.~Lettmann, B.~Linh, J.~Liu, Z.~Liu,
%  C.~Lizarazo, C.~Louchart, R.~Lozeva, K.~Matsui, T.~Miyazaki, K.~Moschner,
%  M.~Nagamine, N.~Nakatsuka, S.~Nishimura, C.~Nita, C.~Nobs, L.~Olivier,
%  S.~Ota, R.~Orlandi, Z.~Patel, P.~Regan, M.~Rudigier, E.~èûahin, T.~Saito,
%  P.A. Sº¢derstrº¢m, I.~Stefan, T.~Sumikama, D.~Suzuki, Z.~Vajta, V.~Vaquero,
%  V.~Werner, K.~Wimmer, J.~Wu, Z.~Xu, 
  Physics Letters B \textbf{773}, 492
  (2017)

\bibitem{Morfouace2015}
P.~Morfouace, S.~Franchoo, K.~Sieja, I.~Matea, L.~Nalpas {\it et al.},
%, M.~Niikura,
%  A.~Sánchez-Benítez, I.~Stefan, M.~Assié, F.~Azaiez, D.~Beaumel, S.~Boissinot,
%  C.~Borcea, R.~Borcea, G.~Burgunder, L.~Cáceres, N.D. Séréville, Z.~Dombrádi,
%  J.~Elseviers, B.~Fernández-Domínguez, A.~Gillibert, S.~Giron, S.~Grévy,
%  F.~Hammache, O.~Kamalou, V.~Lapoux, L.~Lefebvre, A.~Lepailleur, C.~Louchart,
%  G.~Marquinez-Duran, I.~Martel, A.~Matta, D.~Mengoni, D.~Napoli, F.~Recchia,
%  J.A. Scarpaci, D.~Sohler, O.~Sorlin, M.~Stanoiu, C.~Stodel, J.C. Thomas, Z.~Vajta, 
  Phys. Lett. B \textbf{751}, 306  (2015)

\bibitem{Giron}
S.~Giron, F.~Hammache, N.~de~S\'er\'eville, P.~Roussel, J.~Burgunder {\it et al.},
%  M.~Moukaddam, D.~Beaumel, L.~Caceres, G.~Duch\^ene, E.~Cl\'ement,
%  B.~Fernandez-Dominguez, F.~Flavigny, G.~de~France, S.~Franchoo,
%  D.~Galaviz-Redondo, L.~Gasques, J.~Gibelin, A.~Gillibert, S.~Grevy,
%  J.~Guillot, M.~Heil, J.~Kiener, V.~Lapoux, F.~Mar\'echal, A.~Matta, I.~Matea,
%  L.~Nalpas, J.~Pancin, L.~Perrot, A.~Obertelli, R.~Raabe, J.A. Scarpaci,
%  K.~Sieja, O.~Sorlin, I.~Stefan, C.~Stodel, M.~Takechi, J.C. Thomas,
%  Y.~Togano, 
  Phys. Rev. C \textbf{95}, 035806 (2017)

\bibitem{Morfouace2016}
P.~Morfouace, S.~Franchoo, K.~Sieja, I.~Stefan, N.~de~S\'er\'eville {\it et al.},
%  F.~Hammache, M.~Assi\'e, F.~Azaiez, C.~Borcea, R.~Borcea, L.~Grassi,
%  J.~Guillot, B.~Le~Crom, L.~Lefebvre, I.~Matea, D.~Mengoni, D.~Napoli,
%  C.~Petrone, M.~Stanoiu, D.~Suzuki, D.~Testov, 
Phys. Rev. C \textbf{93},  064308 (2016)

\bibitem{Elekes}
Z.~Elekes, A.~Kripk\'o, D.~Sohler, K.~Sieja, K.~Ogata {\it et al.},
%, K.~Yoshida,
%  P.~Doornenbal, A.~Obertelli, G.~Authelet, H.~Baba, D.~Calvet, F.~Ch\^ateau,
%  A.~Corsi, A.~Delbart, J.M. Gheller, A.~Gillibert, T.~Isobe, V.~Lapoux,
%  M.~Matsushita, S.~Momiyama, T.~Motobayashi, H.~Otsu, C.~P\'eron, A.~Peyaud,
%  E.C. Pollacco, J.Y. Rouss\'e, H.~Sakurai, C.~Santamaria, Y.~Shiga,
%  S.~Takeuchi, R.~Taniuchi, T.~Uesaka, H.~Wang, K.~Yoneda, F.~Browne, L.X.
%  Chung, Z.~Dombr\'adi, F.~Flavigny, S.~Franchoo, F.~Giacoppo, A.~Gottardo,
%  K.~Hady\ifmmode \acute{n}\else \'{n}\fi{}ska-Kl\ifmmode~\mbox{\k{e}}\else
%  \k{e}\fi{}k, Z.~Korkulu, S.~Koyama, Y.~Kubota, J.~Lee, M.~Lettmann,
%  C.~Louchart, R.~Lozeva, K.~Matsui, T.~Miyazaki, M.~Niikura, S.~Nishimura,
%  L.~Olivier, S.~Ota, Z.~Patel, E.~Sahin, C.~Shand, P.A. S\"oderstr\"om,
%  I.~Stefan, D.~Steppenbeck, T.~Sumikama, D.~Suzuki, Z.~Vajta, V.~Werner,
%  J.~Wu, Z.~Xu, 
Phys. Rev. C \textbf{99}, 014312 (2019)

\bibitem{Orlandi}
R.~Orlandi, et~al., Physics Letters B \textbf{740}, 298 (2015)

\bibitem{N3LO}
E.~Epelbaum, H.W. Hammer, U.G. Mei\ss{}ner, Rev. Mod. Phys. \textbf{81}, 1773
  (2009)

\bibitem{vlowk}
S.~Bogner, T.~Kuo, A.~Schwenk, Phys. Rep. \textbf{386}, 1 (2003)

\bibitem{Litzinger}
J.~Litzinger, et~al., Phys. Rev. C \textbf{92}, 064322 (2015)

\bibitem{Czerwinski-Br}
M.~Czerwi\ifmmode~\acute{n}\else \'{n}\fi{}ski, T.~Rzaca-Urban, W.~Urban,
  P.~Baczyk, K.~Sieja {\it et al.},
%  , B.M. Nyak\'o, J.~Tim\'ar, I.~Kuti, T.G. Tornyi,
%  L.~Atanasova, A.~Blanc, M.~Jentschel, P.~Mutti, U.~K\"oster, T.~Soldner,
%  G.~de~France, G.S. Simpson, C.A. Ur, 
Phys. Rev. C \textbf{92}, 014328 (2015)

\bibitem{Zn80}
J.~Van~de Walle, F.~Aksouh, T.~Behrens, V.~Bildstein, A.~Blazhev {\it et al.},
%  J.~Cederk\"all, E.~Cl\'ement, T.E. Cocolios, T.~Davinson, P.~Delahaye,
%  J.~Eberth, A.~Ekstr\"om, D.V. Fedorov, V.N. Fedosseev, L.M. Fraile,
%  S.~Franchoo, R.~Gernhauser, G.~Georgiev, D.~Habs, K.~Heyde, G.~Huber,
%  M.~Huyse, F.~Ibrahim, O.~Ivanov, J.~Iwanicki, J.~Jolie, O.~Kester,
%  U.~K\"oster, T.~Kr\"oll, R.~Kr\"ucken, M.~Lauer, A.F. Lisetskiy, R.~Lutter,
%  B.A. Marsh, P.~Mayet, O.~Niedermaier, M.~Pantea, R.~Raabe, P.~Reiter,
%  M.~Sawicka, H.~Scheit, G.~Schrieder, D.~Schwalm, M.D. Seliverstov, T.~Sieber,
%  G.~Sletten, N.~Smirnova, M.~Stanoiu, I.~Stefanescu, J.C. Thomas, J.J.
%  Valiente-Dob\'on, P.V. Duppen, D.~Verney, D.~Voulot, N.~Warr, D.~Weisshaar,
%  F.~Wenander, B.H. Wolf, M.~Zieli\ifmmode~\acute{n}\else \'{n}\fi{}ska, 
Phys. Rev. C \textbf{79}, 014309 (2009)

\bibitem{NNDC}
http://www.nndc.bnl.gov/

\bibitem{Ni78-nature}
R.~Taniuchi, C.~Santamaria, P.~Doornenbal, et~al., Nature \textbf{569}, 53
  (2019)

\bibitem{sieja-Zr}
K.~Sieja, F.~Nowacki, K.~Langanke, G.~Martinez-Pinedo, Phys. Rev. \textbf{C79},
  064310 (2009).
\newblock \doi{10.1103/PhysRevC.79.064310}

\bibitem{Urban-y}
W.~Urban, K.~Sieja, G.S. Simpson, H.~Faust, T.~Rzaca-Urban {\it et al.},
% A.~Z\l{}omaniec,
%  M.~\L{}ukasiewicz, A.G. Smith, J.L. Durell, J.F. Smith, B.J. Varley,
%  F.~Nowacki, I.~Ahmad, 
Phys. Rev. C \textbf{79}, 044304 (2009).
\newblock \doi{10.1103/PhysRevC.79.044304}

\bibitem{Urban-Sr}
T.~Rzaca-Urban, K.~Sieja, W.~Urban, F.~Nowacki, J.L. Durell {\it et al.},
%, A.G. Smith,
%  I.~Ahmad, 
Phys. Rev. C \textbf{79}, 024319 (2009).
\newblock \doi{10.1103/PhysRevC.79.024319}

\bibitem{Simpson-rb}
G.S. Simpson, W.~Urban, K.~Sieja, J.A. Dare, J.~Jolie {\it et al.},
%, A.~Linneman, R.~Orlandi,
%  A.~Scherillo, A.G. Smith, T.~Soldner, I.~Tsekhanovich, B.J. Varley,
%  A.~Z\l{}omaniec, J.L. Durell, J.F. Smith, T.~Rzaca-Urban, H.~Faust, I.~Ahmad,
%  J.P. Greene, 
Phys. Rev. C \textbf{82}, 024302 (2010).
\newblock \doi{10.1103/PhysRevC.82.024302}

\bibitem{urban-rb}
W.~Urban, K.~Sieja, G.S. Simpson, T.~Soldner, T.~Rzaca-Urban {\it et al.},
%, A.~Z\l{}omaniec,
%  I.~Tsekhanovich, J.A. Dare, A.G. Smith, J.L. Durell, J.F. Smith, R.~Orlandi,
%  A.~Scherillo, I.~Ahmad, J.P. Greene, J.~Jolie, A.~Linneman, 
Phys. Rev. C  \textbf{85}, 014329 (2012).
\newblock \doi{10.1103/PhysRevC.85.014329}

\bibitem{Czerwinski2019}
J.~Wisniewski, W.~Urban, M.~Czerwinski, J.~Kurpeta, A.~P\l{}ochocki {\it et al.},
%  M.~Pomorski, T.~Rz\k{a}ca-Urban, K.~Sieja, L.~Canete, T.~Eronen, S.~Geldhof,
%  A.~Jokinen, A.~Kankainen, I.D. Moore, D.A. Nesterenko, H.~Penttil\"a,
%  I.~Pohjalainen, S.~Rinta-Antila, A.~de~Roubin, M.~Vil\'en, 
Phys. Rev. C  \textbf{100}, 054331 (2019)

\bibitem{Materna2015}
T.~Materna, W.~Urban, K.~Sieja, U.~K\"oster, H.~Faust {\it et al.},
%  M.~Czerwi\ifmmode~\acute{n}\else \'{n}\fi{}ski, T.~Rz\k{a}ca-Urban,
%  C.~Bernards, C.~Fransen, J.~Jolie, J.M. Regis, T.~Thomas, N.~Warr, 
Phys. Rev.  C \textbf{92}, 034305 (2015)

\bibitem{Didierjean}
F.~Didierjean, D.~Verney, G.~Duch\^ene, J.~Litzinger, K.~Sieja {\it et al.},
%, A.~Dewald,
%  A.~Goasduff, R.~Lozeva, C.~Fransen, G.~de~Angelis, S.~Aydin, D.~Bazzacco,
%  A.~Bracco, S.~Bottoni, L.~Corradi, F.~Crespi, E.~Ellinger, E.~Farnea,
%  E.~Fioretto, S.~Franchoo, A.~Gottardo, L.~Grocutt, M.~Hackstein, F.~Ibrahim,
%  K.~Kolos, S.~Leoni, S.~Lenzi, S.~Lunardi, R.~Menegazzo, D.~Mengoni,
%  C.~Michelagnoli, T.~Mijatovic, V.~Modamio, O.~M\"oller, G.~Montagnoli,
%  D.~Montanari, A.I. Morales, D.~Napoli, M.~Niikura, F.~Recchia, E.~Sahin,
%  F.~Scarlassara, L.~Sengele, S.~Szilner, J.F. Smith, A.M. Stefanini, C.~Ur,
%  J.J. Valiente-Dob\'on, V.~Vandone, 
Phys. Rev. C \textbf{96}, 044320 (2017)

\bibitem{Kolos2013}
K.~Kolos, D.~Verney, F.~Ibrahim, F.~Le~Blanc, S.~Franchoo {\it et al.},
%, K.~Sieja,
%  F.~Nowacki, C.~Bonnin, M.~Cheikh~Mhamed, P.V. Cuong, F.~Didierjean,
%  G.~Duch\^ene, S.~Essabaa, G.~Germogli, L.H. Khiem, C.~Lau, I.~Matea,
%  M.~Niikura, B.~Roussi\`ere, I.~Stefan, D.~Testov, J.C. Thomas, 
Phys. Rev. C  \textbf{88}, 047301 (2013)

\bibitem{Lettman}
M.~Lettmann, V.~Werner, N.~Pietralla, P.~Doornenbal, A.~Obertelli {\it et al.},
%, T.R.
%  Rodr\'{\i}guez, K.~Sieja, G.~Authelet, H.~Baba, D.~Calvet, F.~Ch\^ateau,
%  S.~Chen, A.~Corsi, A.~Delbart, J.M. Gheller, A.~Giganon, A.~Gillibert,
%  V.~Lapoux, T.~Motobayashi, M.~Niikura, N.~Paul, J.Y. Rouss\'e, H.~Sakurai,
%  C.~Santamaria, D.~Steppenbeck, R.~Taniuchi, T.~Uesaka, T.~Ando, T.~Arici,
%  A.~Blazhev, F.~Browne, A.~Bruce, R.J. Caroll, L.X. Chung, M.L. Cort\'es,
%  M.~Dewald, B.~Ding, F.~Flavigny, S.~Franchoo, M.~G\'orska, A.~Gottardo,
%  A.~Jungclaus, J.~Lee, B.D. Linh, J.~Liu, Z.~Liu, C.~Lizarazo, S.~Momiyama,
%  K.~Moschner, S.~Nagamine, N.~Nakatsuka, C.~Nita, C.R. Nobs, L.~Olivier,
%  Z.~Patel, Z.~Podoly\'ak, M.~Rudigier, T.~Saito, C.~Shand, P.A.
%  S\"oderstr\"om, I.~Stefan, V.~Vaquero, K.~Wimmer, Z.~Xu, 
Phys. Rev. C  \textbf{96}, 011301 (2017)

\bibitem{Se87}
T.~Rzaca-Urban, M.~Czerwi\ifmmode~\acute{n}\else \'{n}\fi{}ski, W.~Urban, A.G.
  Smith, I.~Ahmad {\it et al.},
%  , F.~Nowacki, K.~Sieja, 
  Phys. Rev. C \textbf{88}, 034302 (2013)

\bibitem{Gratchev}
I.~Gratchev, et~al., Phys. Rev. C \textbf{95}(5), 051302 (2017)

\bibitem{sieja2013}
K.~Sieja, T.R. Rodr\'iguez, K.~Kolos, D.~Verney, Phys. Rev. C \textbf{88},
  034327 (2013)

\bibitem{Fisker2001}
J.~Fisker, et~al., Atomic Data and Nuclear Data Tables \textbf{79}, 241 (2001)

\bibitem{Rochman17}
D.~Rochman, S. Goriely, A.J. Koning,  H. Ferroukhi, Phys. Lett. B \textbf{764}, 109 (2017)

\bibitem{TALYS}
A.J. Koning, D. Rochman, Nucl. Data Sheets \textbf{113}, 2841 (2012); {\it http://www.talys.eu}

\bibitem{czerwinski}
M.~Czerwinski, T.~Rzaca-Urban, K.~Sieja, H.~Sliwinska, W.~Urban {\it et al.},
%, A.G. Smith,
%  J.F. Smith, G.S. Simpson, I.~Ahmad, J.P. Greene, T.~Materna, 
Phys. Rev. C  \textbf{88}, 044314 (2013)


\bibitem{Goriely11} S. Goriely, A. Bauswein and H.-T. Janka, Astrophys. J. Lett. \textbf{738}, L32 (2011).

\bibitem{Goriely15} S. Goriely, A. Bauswein, O. Just, and H.-T. Janka, Mon. Not. Roy. Astron. Soc. \textbf{452}, 3894 (2015).

\bibitem{Just15} O. Just, A. Bauswein, R. Ardevol-Pulpillo, S. Goriely and  H.-T. Janka, Mon. Not. Roy. Astron. Soc.  \textbf{448}, 541 (2015).

\bibitem{Abbott17} B. Abbott, R. Abbott, T. Abbott, et al., Phys. Rev. Lett. \textbf{119}, 161101 (2017).

\bibitem{Ardevol19} R. Ardevol-Pulpillo,   H.-T. Janka, O. Just, and A. Bauswein, Mon. Not. Roy. Astron. Soc. \textbf{485}, 4754 (2019).

\bibitem{Lemaitre20} J.-F. Lema\^\i tre, S. Goriely, A. Bauswein and H.-T. Janka, Astrophys. Phys, Rev. C (2020) submitted.

\bibitem{Goriely99} S. Goriely, Astron. Astrophys. \textbf{342}, 881 (1999).


\end{thebibliography}
%\begin{thebibliography}{999}
%
%\providecommand{\url}[1]{{#1}}
%\providecommand{\urlprefix}{URL }
%\expandafter\ifx\csname urlstyle\endcsname\relax
%  \providecommand{\doi}[1]{DOI \discretionary{}{}{}#1}\else
%  \providecommand{\doi}{DOI \discretionary{}{}{}\begingroup
%  \urlstyle{rm}\Url}\fi

 \end{document}